\documentclass[english,aps,pra,twocolumn,superscriptaddress,showpacs]{revtex4}
\usepackage{amsmath}
\usepackage[T1]{fontenc}
\usepackage[latin1]{inputenc}
\usepackage{amsmath,amsfonts}
\usepackage{epsf}
\usepackage{graphicx}
\usepackage{amssymb}
\usepackage{babel}
\usepackage{color}
\usepackage{ifpdf}
\usepackage{comment}

\makeatletter

\begin{document}

\title{Significance of dressed molecules in a quasi-two-dimensional Fermi gas with spin-orbit coupling}

\author{Ren Zhang}
\affiliation{Department of Physics, Renmin University of China, Beijing 100872, People's Republic of China}
\author{Fan Wu}
\affiliation{Key Laboratory of Quantum Information, University of Science and Technology of China, CAS, Hefei, Anhui, 230026, People's Republic of China}
\author{Jun-Rong Tang}
\affiliation{Department of Physics, Renmin University of China, Beijing 100872, People's Republic of China}
\author{Guang-Can Guo}
\affiliation{Key Laboratory of Quantum Information, University of Science and Technology of China, CAS, Hefei, Anhui, 230026, People's Republic of China}
\author{Wei Yi}
\email{wyiz@ustc.edu.cn}
\affiliation{Key Laboratory of Quantum Information, University of Science and Technology of China, CAS, Hefei, Anhui, 230026, People's Republic of China}
\author{Wei Zhang}
\email{wzhangl@ruc.edu.cn}
\affiliation{Department of Physics, Renmin University of China, Beijing 100872, People's Republic of China}

\begin{abstract}
We investigate the properties of a spin-orbit coupled quasi-two-dimensional Fermi gas with tunable $s$-wave interaction between the two spin species. By analyzing the two-body bound state, we find that the population of the excited states in the tightly-confined axial direction can
be significant when the two-body binding energy becomes comparable to or exceeds the axial confinement. Since the Rashba spin-orbit coupling that we study here
tends to enhance the two-body binding energy, this effect can become prominent at unitarity
or even on the BCS side of the Feshbach resonance.
To study the impact of these excited modes along the third dimension, we adopt an effective
two-dimensional Hamiltonian in the form of a two-channel model, where the dressed molecules in the closed channel consist of the conventional Feshbach molecules as well as the excited states occupation in the axial direction.
With properly renormalized interactions between atoms and dressed molecules, we find that both the density distribution and the phase structure in the trap can be significantly modified near a wide Feshbach resonance. In particular, the stability region of the topological superfluid phase is increased. Our findings provide a proper description for a
quasi-two-dimensional Fermi gas under spin-orbit coupling,
and are helpful for the experimental search for the topological superfluid phase in ultra-cold Fermi gases.

%Our findings are helpful for the experimental search for the topological superfluid phase in ultra-cold Fermi gases, and have interesting %implications for quasi-low-dimensional polarized Fermi gases in general.
\end{abstract}
\maketitle
%%%%%%%%%%%%%%
%%%%%%%%%%%%%%
\section{Introduction}

%Topological superfluid (TSF) phase is an unconventional superfluid state with superfluid gap
%in the bulk and topologically protected gapless modes at boundaries. As Majorana zero modes can
%emerge at the core of vortex excitations above the TSF state, these systems have attracted a considerable
%amount of attention recently~\cite{nayak, moore, dassarma, gurarie, tewari}.
%There have been various proposals for realizing the TSF state in condensed matter systems.
%In particular, J. Sau et al. suggested that such a state can be realized at the interface of
%semiconductor/superconductor heterostructures with spin-orbit coupling (SOC), $s$-wave pairing superfluidity
%and an external Zeeman field~\cite{sau}. Notably, these key elements are also available in the context of
%quasi-two-dimensional (quasi-2D) ultra-cold Fermi gases near an $s$-wave Feshbach resonance. Importantly,
%with recent experimental realizations of Abelian gauge field~\cite{lin-exp1} and SOC in ultra-cold atoms~\cite{lin-exp2, zhang-exp, mit-exp},
%it seems plausible that Majorana fermions can be realized and manipulated in ultra-cold Fermi gases.

Recent experimental realizations of spin-orbit coupling (SOC) in ultra-cold
atoms~\cite{lin-exp2, zhang-exp, mit-exp} stimulate a new wave of exploration for exotic phases
in polarized Fermi gases~\cite{zhang-sarma, sato, shenoy, gongming, yu, hu, iskin, yi,
sademelo, jzhou, yang, chuanwei, helianyi}.
The competition between polarization, pairing superfluidity and SOC gives rise to
rich phase structures. In three dimensions, two different topologically non-trivial
superfluid phases with gapless excitations exist~\cite{gongming, yu, hu, iskin, yi, sademelo}.
In two dimensions, a topological superfluid (TSF) phase~\cite{sau}
which supports Majorana zero mode at the core of
vortex excitations can be stabilized~\cite{jzhou,yang}.
It has been demonstrated that the topological superfluid phase has large stability region near a wide Feshbach resonance, where the strong interaction also ensures the stability of the system against collisional losses and increases the critical temperature~\cite{chuanwei,helianyi}. This makes the quasi-2D polarized Fermi gas with SOC and strong interaction an ideal system for the realization of TSF state. Experimentally, quasi-2D Fermi gases are implemented by applying a strong confinement along the axial ($z$) direction and a weak harmonic trapping potential in the transverse ($x$-$y$) plane. Depending on the experimental design, it is possible to realize either a one-dimensional array~\cite{ol} or a single layer~\cite{vale} of quasi-2D Fermi gases. When the energy of the two-body bound state is small, the excited states in the axial direction are not significantly occupied at low temperatures. In this case, the axial degrees of freedom can be easily integrated out, leaving an effective two-dimensional (2D) Hamiltonian with renormalized atom-atom interactions~\cite{randeria,petrov,wouters}. However, with large two-body bound state energy, it has been shown that the occupation of axial excited state can be significant,
and the degrees of freedom along the third dimension can lead to measurable many-body effects on the BEC side of the Feshbach resonance~\cite{jason-pra06,Duan-07, wzduan,wzduan2,demler}.
In this case, the many-body properties can be characterized by an effective two-channel model with renormalized interactions between atoms and dressed molecules, where the dressed molecules consist of Feshbach molecules and atoms in the axial excited states.

In this work, we extend the effective two-channel model to describe quasi-2D Fermi gases
with SOC. To incorporate the effective Zeeman field which will be present in certain experimental
schemes~\cite{lin-exp2, zhang-exp, mit-exp}, we also take population imbalance into consideration.
We demonstrate that the occupation of the axial excited states can lead to non-trivial effects and hence
such an extension is important for a proper description of the system. We first solve the two-body
problem within quasi-2D confinement, and find that
the presence of SOC causes an increase of the excited modes population along the
axial direction. This observation indicates that the third dimension would matter in a larger
parameter region across the BCS-BEC crossover, and have to be taken into account to describe
the quasi-2D, but truly three-dimensional (3D) system. We then adopt an effective 2D Hamiltonian in a two-channel model
form and determine the parameters by matching single- and two-body physics of the original Hamiltonian.
In order to demonstrate the significance of dressed molecules, we investigate the phase structure and
density distribution of a trapped quasi-2D Fermi gas and compare results of our proposed
two-channel model and those of the oversimplified single-channel model.
We find that by correctly incorporating the dressed molecules, both the phase structure and
density profiles can be significantly modified near a Feshbach resonance.
Typically, the phase boundaries between the TSF state and the conventional superfluid (SF)
state are shifted in the trap. Under certain conditions, the inclusion of dressed molecules can even
qualitatively alter the in-trap phase structure and induce an additional TSF phase which is not
present in a single-channel calculation. In either case, the stability region of the TSF phase is increased.
Interestingly, we find that even a small closed-channel population can lead to significant changes in
density profiles and/or phase structures. This suggests that the significance of dressed molecules
may be probed experimentally near a wide Feshbach resonance.

In the framework of the effective two-channel model, the dressed molecules are modelled as structureless
bosons, which consist of equal number of spin-up and spin-down fermions. This is in contrast 
to the various exotic superfluid phases. Indeed,
the effects of dressed molecules on the phase structure can be qualitatively understood by considering
the accommodation of spin imbalance among various phases in the trap. For a 3D polarized Fermi gas without SOC, typically, the competition between pairing and polarization can lead to the emergence
of exotic phases, e.g., Fulde-Ferrel-Larkin-Ovchinnikov (FFLO) phase~\cite{fflo1,fflo2} or the
breached pairing (BP) phase~\cite{bp1} etc., in addition to the quantum phase transition between
the SF and the normal states. In fact, as the conventional $s$-wave pairing superfluid state does
not support population imbalance at zero temperature, normal phase or more exotic phases
must exist in the trap to accommodate polarization. The requirement of accommodating population
imbalance is more crucial in a quasi-low-dimensional configuration, in which case atoms are
inevitably populated to excited states in the strongly confined direction. These axial
excited states are induced by inter-species $s$-wave interaction, such that they are significantly populated
only when a spin-up fermion and a spin-down fermion form a tightly bound molecule.
Thus, there is no spin imbalance in the axially excited states, and the spin imbalance has to be taken care of by the atoms left in the ground state of the confinement. As a consequence, we note that the inclusion of dressed molecules in a
polarized quasi-2D Fermi gas without SOC would lead to measurable many-body effects,
including a possible enhanced stability region of the exotic phases there.

The situation is similar in the presence of SOC. Here, pairing parity is mixed,
which leads to a small amount of population imbalance in the SF phase in the presence of an effective Zeeman field at zero temperature.
However, this is overcome by the absence of normal state due to an SOC-induced pairing instability. More importantly,
when the Zeeman splitting induced by the effective magnetic field is much smaller than the
axial confinement, the spin imbalance of the axial excited states becomes negligible.  
Hence, the lack of mechanism of the system to accommodate polarization otherwise leads to an
enlarged stability region of the exotic TSF phase in a trap.
With experimental achievements in preparing and probing quasi-low-dimensional Fermi
gases~\cite{2dgasexp1,2dgasexp2,2dgasexp3,2dgasexp4}, as well as the recent realization of SOC in
ultra-cold Fermi gases~\cite{zhang-exp,mit-exp}, our findings here should be helpful for future experiments.

The paper is organized as follows: in Sec. II, we consider a two-body problem in quasi two dimensions
in the presence of SOC and solve for the two-body bound state. We find that the
population in the excited harmonic modes along the axial direction can become significant
at unitarity or even on the BCS side of the resonance. To incorporate the effect of these excited modes,
in Sec. III we write down an effective 2D Hamiltonian in a two-channel model form, and fix the parameters
by matching single- and two-body physics. We then study typical phase structures of a trapped gas
by using the effective two-channel model and discuss the effects of dressed molecules in Sec. IV.
Finally, we summarize in Sec. V.
%%%%%%%%%
\section{Two-body bound state in quasi-2D confinement}
\label{sec:boundstate}

We consider a two-component Fermi gas confined in a one-dimensional
harmonic potential with trapping frequency $\omega_z$. In the presence of SOC,
the system can be described by a conventional two-channel field theory around
a Feshbach resonance~\cite{holland-01, timmermans-99}
\begin{eqnarray}
\label{eqn:real-H}
H = H_0 + H_{\rm soc} + H_{\rm bf} + H_{\rm int}.
\end{eqnarray}
The terms in the Hamiltonian are
\begin{eqnarray}
\label{eqn:real-H-2}
H_0 &=& \sum_{\sigma = \uparrow, \downarrow} \int d^3 \boldsymbol{r} \psi_{\sigma}^\dagger
\left( - \frac{\hbar^2 \nabla^2}{2m_f} + \frac{1}{2} m_f \omega_z^2 z^2 \right) \psi_{\sigma}
\nonumber \\
&&
+ \int d^3\boldsymbol{r} \phi^\dagger
\left( - \frac{\hbar^2 \nabla^2}{4m_f} + m_f \omega_z^2 z^2 + {\bar \nu}_b \right) \phi,
\nonumber \\
H_{\rm bf} &=& {\bar g}_b \int d^3\boldsymbol{r}
\left( \phi^\dagger \psi_\downarrow \psi_\uparrow + {\rm H.C.} \right)
\nonumber \\
H_{\rm int} &=&
{\bar U}_b \int d^3\boldsymbol{r} \psi_\uparrow^\dagger \psi_\downarrow^\dagger
\psi_\downarrow \psi_\uparrow,
\end{eqnarray}
where $\psi_\sigma(\boldsymbol{r})$ is the fermionic field operator with spin index $\sigma = (\uparrow, \downarrow)$,
$\phi(\boldsymbol{r})$ is the bosonic molecular field operator,
${\bar \nu}_b$ is the bare detuning,
${\bar g}_b$ is the bare atom-molecule coupling constant,
${\bar U}_b$ is the bare background scattering amplitude, and
H.C. stands for Hermitian conjugate.
Here, we assume a contact interaction between atoms.
The bare scattering parameters are related to the physical ones (with subscript $p$)
via the standard renormalization relations \cite{qijinpr}
\begin{eqnarray}
\label{eqn:renorm}
&&U_{c}^{-1} = - \int \frac{d^3\boldsymbol{k}}{(2\pi^3)} \frac{1}{2 {\bar\epsilon}_{\boldsymbol{k}}},
\quad
\Gamma^{-1} = 1 + \frac{{\bar U}_p}{U_c},
\nonumber \\
&&{\bar U}_p = \Gamma^{-1} {\bar U}_b, \quad {\bar g}_p = \Gamma^{-1} {\bar g}_b, \quad
{\bar \nu}_p = {\bar \nu}_b + \Gamma \frac{{\bar g}_p^2}{U_c}.
\end{eqnarray}
Here, ${\bar \epsilon}_{\boldsymbol{k}} = \hbar^2 k^2 / (2m_f)$ is the dispersion relation
for atoms with mass $m_f$ and 3D momentum $\boldsymbol{k}$,
and the integral is taken in three dimensions with an explicit 2D energy cutoff $E_c$,
so $U_c^{-1} = \sqrt{E_c} / 2^{3/2} \pi$.
The physical parameters can be obtained from the scattering measurements with
\begin{eqnarray}
\label{eqn:physical}
{\bar U}_p &=&\frac{4 \pi \hbar^2 a_{\rm bg}}{m_f},
\hspace{0.2cm}{\bar g}_p = \sqrt{\frac{4 \pi \hbar^2 \mu_{\rm co} W |a_{\rm bg}|}{m_f}},\nonumber\\
{\bar \nu}_p &=& \mu_{\rm co} (B-B_0),
\end{eqnarray}
where $a_{\rm bg}$ is the 3D background scattering length, $W$ is the resonance width,
$\mu_{\rm co}$ is the difference in magnetic moments of the closed and open channels,
and $B_0$ is the resonance position.

The second term in the Hamiltonian Eq. (\ref{eqn:real-H})
represents the spin-orbit coupling (SOC).
In the following discussion, we consider only the Rashba type SOC as
\begin{equation}
H_{\rm soc} = - i \hbar {\bar \lambda} \int d^3 \boldsymbol{r} {\bar \psi}^\dagger
\left( \sigma_x \partial_x + \sigma_y \partial_y \right) {\bar \psi}.
\end{equation}
The parameter ${\bar \lambda}$ is the SOC constant, $\sigma_{i=x,y}$ are the Pauli
matrices, and ${\bar \psi} = ( \psi_\uparrow, \psi_\downarrow )^{T}$. This SOC term couples
particles with different spins such that spin is no longer a good quantum number.
Instead, the single-particle Hamiltonian can be diagonalized in the helicity basis
where each helix corresponds to particles with in-plane spin parallel or antiparallel
to the in-plane momentum.

To discuss the two-body bound state, we notice that
the center-of-mass (CoM) degree of freedom along the axial direction is not affected by the interaction
nor the SOC and can be separated from the relative coordinate under a harmonic potential.
Thus, we can assume the CoM degree of freedom is in the ground
harmonic mode along the $z$-direction.
In this CoM frame, the terms in Hamiltonian Eq. (\ref{eqn:real-H}) can be rewritten
in a second quantized form by expanding the field operators $\psi_{\sigma}$ and
$\phi$ in terms of harmonic oscillators along the $z$-direction and plane waves
in the $x$-$y$ plane
\begin{eqnarray}
\label{eqn:H1}
H_0 &=& \sum_{m, {\bf k}, \sigma}
\left(\varepsilon_{m} + \epsilon_{{\bf k}} \right)
c_{m, {\bf k},\sigma}^\dagger c_{m, {\bf k},\sigma}\nonumber\\
&& \hspace{1cm}
+\sum_{\ell, {{\bf q}}} \left(\nu_b + \varepsilon_{\ell} + \epsilon_{{\bf q}}/2 \right)
b_{\ell,{{\bf q}}}^\dagger b_{\ell,{{\bf q}}},
\nonumber \\
H_{\rm soc} &=& \lambda \sum_{m,{\bf k}}
\left[ (k_x - ik_y) c_{m, {\bf k},\uparrow}^\dagger c_{m, {\bf k},\downarrow}
+ {\rm H.C.} \right],
\nonumber\\
%\end{eqnarray}
%\begin{eqnarray}
H_{\rm bf}  &=&  g_b \sum_{m,n, \ell, {\bf k}, {\bf q}} \gamma_{mn \ell}
\bigg( b_{\ell, {\bf q}}^\dagger c_{m, -{\bf k}+{\bf q}/2, \downarrow}
c_{n, {\bf k}+{\bf q}/2, \uparrow}
\nonumber \\
&& \hspace{2cm}
+ {\rm H.C.} \bigg)\nonumber\\
%\end{eqnarray}
%\begin{eqnarray}
H_{\rm int} &=& U_b \sum_{m,n, {\bf k},m^\prime,n^\prime, {\bf k}^\prime,{\bf q}}
\gamma_{mn}^{m^\prime n^\prime}
c_{m, {\bf k}+{\bf q}/2,\uparrow}^\dagger
c_{n, -{\bf k}+{\bf q}/2,\downarrow}^\dagger\nonumber\\
&& \hspace{1cm}
\times c_{n^\prime, -{\bf k}^\prime+{\bf q}/2,\downarrow}
c_{m^\prime, {\bf k}^\prime+{\bf q}/2,\uparrow}.
\end{eqnarray}
Here, we have used the $z$-direction trapping energy $\hbar \omega_z$ as the energy unit,
and its characteristic length $a_t  = \sqrt{\hbar/(m_f \omega_z)}$ as the length unit.
The corresponding dimensionless parameters are defined as
$g_b = {\bar g}_b a_t^{-3/2} / ({\hbar \omega_z})$,
$U_b = {\bar U}_b a_t^{-3} / ({\hbar \omega_z})$,
$\nu_b = {\bar \nu}_b / ({\hbar \omega_z})$,
and $\lambda = {\bar \lambda} a_t^{-1} / ({\hbar \omega_z})$.
The bosonic field $b_{\ell,{\bf q}}$ represents the molecular state
with axial harmonic mode $\ell$ and 2D transverse momentum ${\bf q}$,
which is also dimensionless in the unit of $a_t^{-1}$.
The fermionic field $c_{m,{\bf k},\sigma}$ represents atomic state
with axial harmonic mode $m$ and 2D momentum ${\bf k}$,
and is characterized with axial mode energy $\varepsilon_m = m+ 1/2$
and plane wave energy $\epsilon_{\bf k} = (k_x^2 + k_y^2)/2$ in the dimensionless form.
The factors appearing in $H_{\rm bf}$ and $H_{\rm int}$ are defined as overlap
of harmonic oscillators
\begin{eqnarray}
\label{eqn-gamma1}
\gamma_{mn\ell} &=& 2^{1/4} \int dz
\Phi_m(z) \Phi_n(z) \Phi_{\ell} (\sqrt{2} z ),
\nonumber \\
\gamma_{mn}^{m^\prime n^\prime} &=&
\int dz  \Phi_m(z) \Phi_n(z)
\Phi_{m^\prime} (z) \Phi_{n^\prime} (z)
\nonumber \\
&=&\sum_{\ell} \gamma_{mn\ell} \gamma_{m^\prime n^\prime \ell}^*.
\end{eqnarray}
Here, $\Phi_n$ is the wavefunction of the $n^{\rm th}$ harmonic oscillator
\begin{eqnarray}
\Phi_n (x) = \frac{e^{-x^2/2}}{\pi^{1/4} \sqrt{2^n n!}} H_n(x)
\end{eqnarray}
with $H_n(x)$ the $n^{\rm th}$ Hermite polynomial.

A general two-body state involving atoms and molecule can be written as the
following ansatz
\begin{eqnarray}
\vert \Psi \rangle_{\ell,{\bf q}} &=&
\bigg( \beta_{\ell,{\bf q}} b_{\ell,{\bf q}}^\dagger +
\nonumber \\
&& \hspace{-1.5cm}
\sum_{m,n,{\bf k}}{}^\prime
\sum_{\sigma, \sigma^\prime}
\eta_{\ell, m,n,{\bf k},{\bf q}}^{\sigma \sigma^\prime}
c_{m, {\bf k}+{\bf q}/2,\sigma}^\dagger
c_{n, -{\bf k}+{\bf q}/2,\sigma^\prime}^\dagger
\bigg) \vert 0 \rangle,
\end{eqnarray}
where $\sum_{m,n,{\bf k}}^\prime$ indicates summation over mode $(m,n)$
and transverse momentum ${\bf k}$ for $k_y>0$. The summation
over spin runs over all four combinations of $(\sigma, \sigma^\prime)$.
The coefficients $\beta_{\ell,{\bf q}}$ and
$\eta_{\ell, m,n,{\bf k},{\bf q}}^{\sigma \sigma^\prime}$ are determined by
solving the Schr{\"o}dinger's equation
$H \vert \Psi \rangle_{\ell,{\bf q}} = E_{\ell,{\bf q}} \vert \Psi \rangle_{\ell,{\bf q}}$
under the normalization relation.

Under the quasi-2D condition, the $z$ confinement is much
greater than the Fermi energy $E_F$ and temperature $k_B T$. In this case,
the population of two-body states with nonzero CoM axial mode $\ell >0$ is very limited,
and the system's properties will be dominated by two-body physics in
the $\ell = 0$ subspace. Next, we focus on two-body states with zero axial
mode $\ell = 0$, and obtain the equations determining the two-body bound state energy
and the corresponding coefficients $\beta$ and $\eta$'s
\begin{widetext}
\begin{eqnarray}
S_p (E_{\bf q}) &=& \left[ U_p - \frac{g_p^2}{\nu_p +\epsilon_{\bf q}/2 - E_{\bf q} }\right]^{-1},
\label{eqn:bindEeqn}
\\
\beta_{\bf q} &=& \left[ 1 - Z_p^2(E_{\bf q}) \frac{\partial S_p(E_{\bf q})}{\partial E_{\bf q}}\right]^{-1/2},
\label{eqn:betaeqn}
\\
|\eta_{m,n,{\bf k}, {\bf q}}^{\uparrow\uparrow}|^2
&=&
|\eta_{m,n,{\bf k}, {\bf q}}^{\downarrow\downarrow}|^2
=
\frac{\lambda^2 |\eta_{m,n,{\bf k}, {\bf q}}^{\rm s}|^2}
{{\cal E}_{m,n,{\bf k}, {\bf q}}^2}
\left[ \left(
k_x - \frac{\lambda^2 q_y (-k_x q_y + k_y q_x)}
{{\cal E}_{m,n,{\bf k}, {\bf q}}^2 - \lambda^2 q^2} \right)^2
+
\left( x \leftrightarrow y \right)
\right],
\label{eqn:etaupup}
\\
\eta_{m,n,{\bf k}, {\bf q}}^{\uparrow\downarrow} &=&
\frac{1}{2} \left( \eta_{m,n,{\bf k}, {\bf q}}^{\rm s}
+
\eta_{m,n,{\bf k}, {\bf q}}^{\rm t}
\right),
\label{eqn:etaupdown} \\
\eta_{m,n,{\bf k}, {\bf q}}^{\downarrow\uparrow} &=&
\frac{1}{2} \left( \eta_{m,n,{\bf k}, {\bf q}}^{\rm t}
-
\eta_{m,n,{\bf k}, {\bf q}}^{\rm s}
\right),
\label{eqn:etadownup}
\end{eqnarray}
where the functions are defined as
\begin{eqnarray}
S_p(E_{\bf q}) & \equiv & \sum_{m,n,{\bf k}} \gamma_{mn0}^2
\left[
{\cal P}_{m,n, {\bf k}, {\bf q}}^{-1}
%\frac{1}{ {\cal E}_{m,n, \boldsymbol{k}, {\boldsymbol{q}}}
%- \frac{4 \lambda^2 k^2}{{\cal E}_{m,n, \boldsymbol{k},{\boldsymbol{q}}}}
%- \frac{16 \lambda^4 \left( k_x q_y - k_y q_x \right)^2}
%{{\cal E}_{m,n, \boldsymbol{k},{\boldsymbol{q}}} \left( 4{\cal E}_{m,n, \boldsymbol{k},{\boldsymbol{q}}}^2 - 4 \lambda^2 q^2 \right)}}
+
\frac{1}{2\epsilon_{\bf k}}
\right],
\label{eqn:sfunc}\\
Z_p(E_{\bf q}) &\equiv & g_p - \frac{U_p}{g_p} \left( \nu_p + \epsilon_{\bf q}/2 - E_{\bf q} \right),
\label{eqn:Zfunc}\\
{\cal P}_{m,n, {\bf k}, {\bf q}}
&\equiv&
{\cal E}_{m,n, {\bf k}, {\bf q}}
- \frac{4 \lambda^2 k^2}{{\cal E}_{m,n, {\bf k},{\bf q}}}
- \frac{4 \lambda^4 \left( k_x q_y - k_y q_x \right)^2}
{{\cal E}_{m,n, {\bf k},{\bf q}} \left( {\cal E}_{m,n, {\bf k},{\bf q}}^2 - \lambda^2 q^2 \right)},
\label{eqn:pfunc}\\
\eta_{m,n,{\bf k},{\bf q}}^{\rm s}
&=&
\frac{2 \gamma_{mn0} Z_p }{{\cal P}_{m,n,{\bf k},{\bf q}}} \beta_{\bf q},
\label{eqn:etaseqn}
\\
\eta_{m,n,{\bf k},{\bf q}}^{\rm t}
&=&
\frac{2 i \lambda^2 \left( - k_x q_y + k_y q_x \right)}
{{\cal E}_{m,n,{\bf k},{\bf q}} - \lambda^2 q^2}
\eta_{m,n,{\bf k},{\bf q}}^{\rm s}
\label{eqn:etateqn}
\end{eqnarray}
\end{widetext}
with ${\cal E}_{m,n, {\bf k},{\bf q}} = E_{\bf q} - k^2 - 1-m-n-q^2/4$.
Here, we drop  the subscript $\ell = 0$ in $E_{\bf q}$, $\beta_{\bf q}$ and $\eta$'s to simplify notation.
The dimensionless physical parameters are related to the bare ones via the
same renormalization relations as in Eq. (\ref{eqn:renorm}).
The coefficients $\gamma_{m n 0} = 0$ for $m+n$ is odd, and
\begin{eqnarray}
\gamma_{mn0} = \frac{(-1)^{(m-n)/2}}{(2 \pi^3)^{1/4} \sqrt{m! n!}}
\Gamma \left( \frac{m+n+1}{2} \right)
\end{eqnarray}
for $m+n$ is even, where $\Gamma(x)$ is the Euler Gamma function.
\begin{figure}[tbp]
\centering
\includegraphics[width=4.5cm]{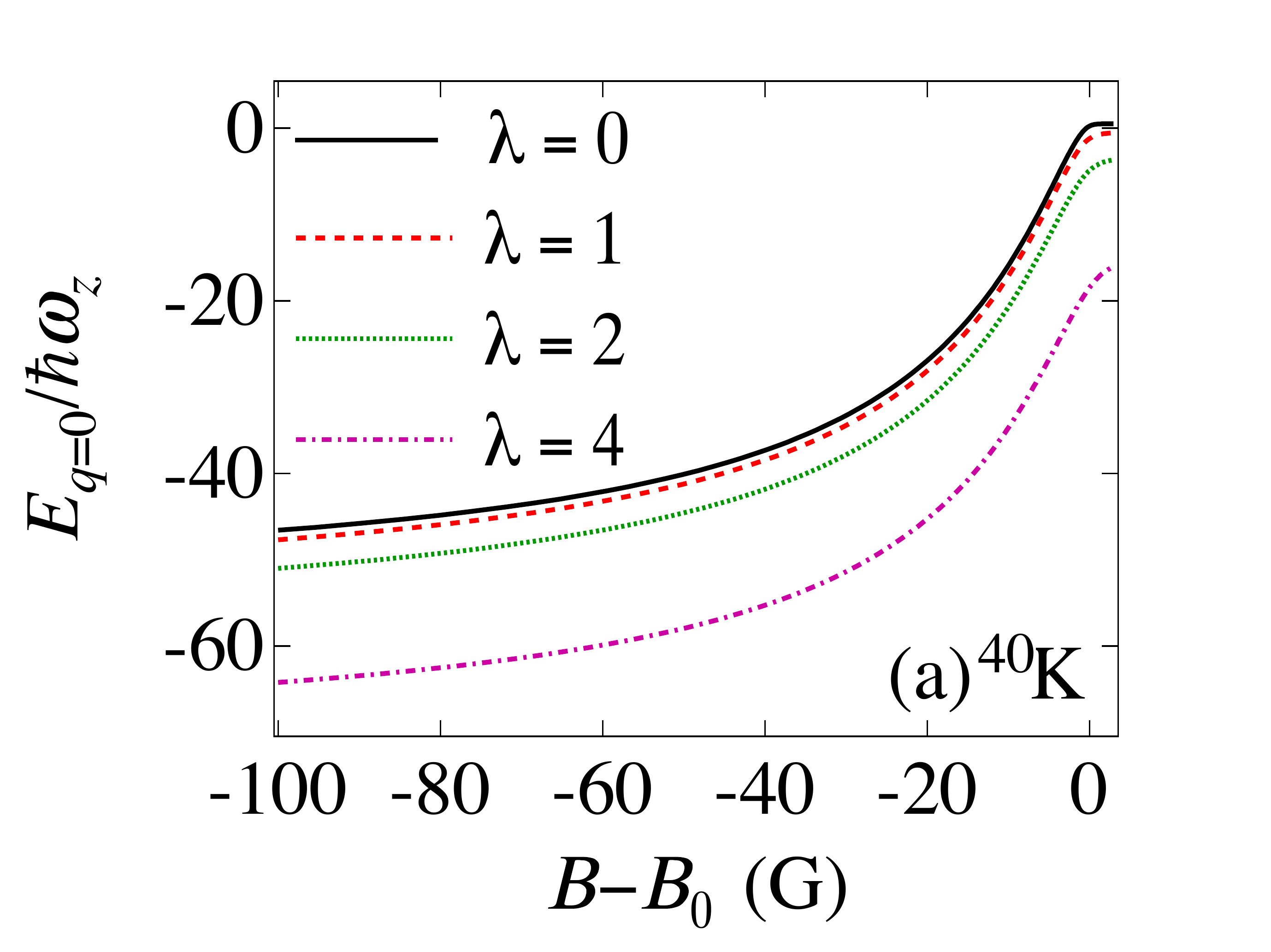}
\hskip-0.5cm
\includegraphics[width=4.5cm]{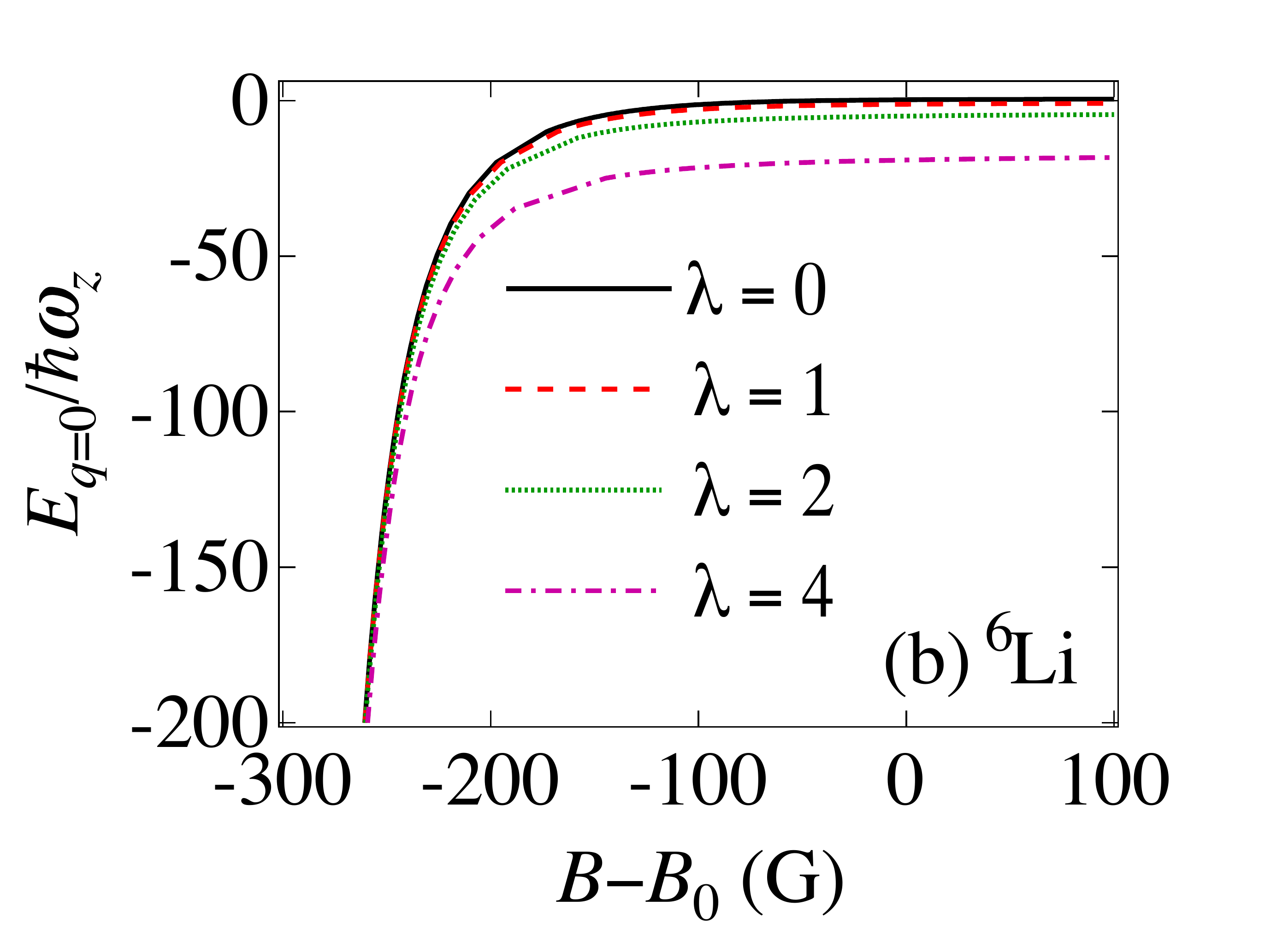}
\includegraphics[width=4.5cm]{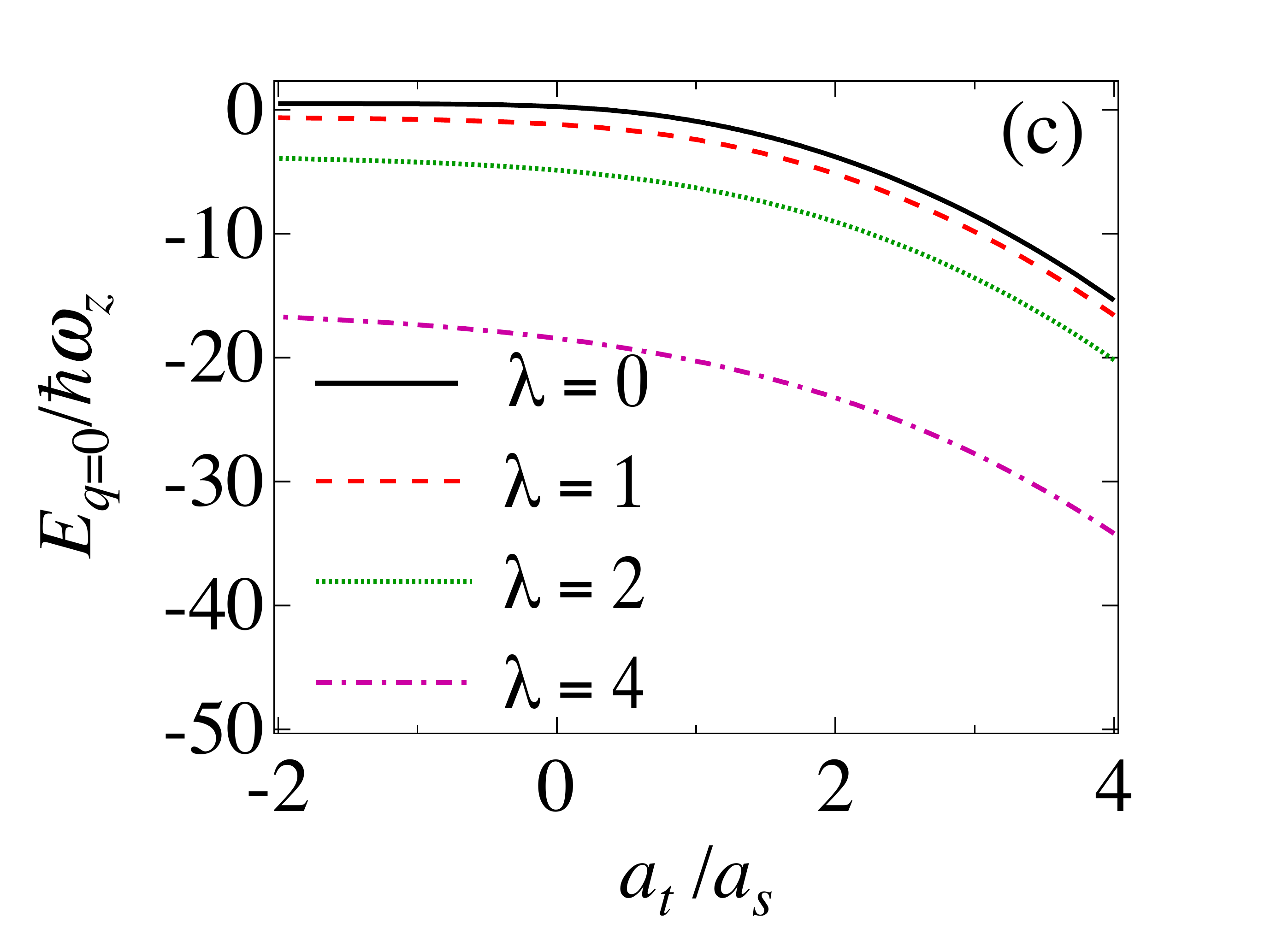}
\hskip-0.5cm
\includegraphics[width=4.5cm]{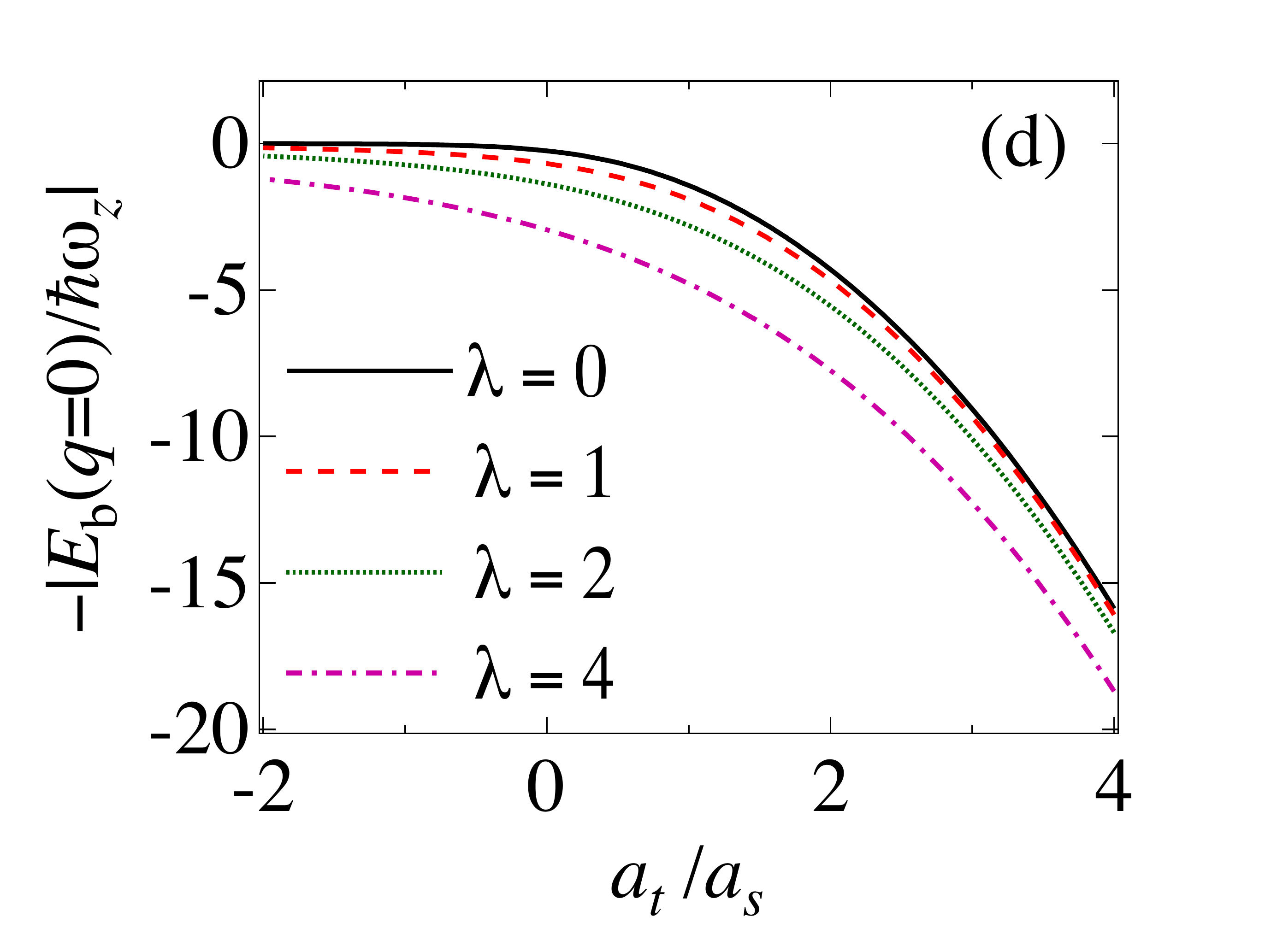}
\caption{(Color online) Two-body bound state energy vs detuning for (a) $^{40}$K and (b) $^{6}$Li
with transverse CoM momentum $q=0$.
(c) When plotted as functions of $a_t/a_s$, results for $^{40}$K and $^{6}$Li are almost indistinguishable,
showing universal behavior around unitarity. (d) The presence of SOC tends
to enhance the two-body binding energy, which is the difference between bound state
energy and open-channel threshold.}
\label{fig:Eb}
\end{figure}

The energy of two-body bound state is determined by solving Eq. (\ref{eqn:bindEeqn}),
and the corresponding eigenstate can be extracted from Eqs. (\ref{eqn:betaeqn} - \ref{eqn:etadownup}).
As typical examples, we focus on the $s$-wave wide Feshbach resonances around $B_0 = 202$ G for $^{40}$K
and $B_0 = 834$ G for $^{6}$Li. The scattering parameters are taken to be the same as the
cases {\it without} SOC, with $W = 8$ G, $a_{bg} = 174 a_B$, $\mu_{\rm co} = 1.68 \mu_B$
for $^{40}$K~\cite{kresonance1,kresonance2,kresonance3},
and $W = 300$ G, $a_{bg} = -1405 a_B$, $\mu_{\rm co} = 2 \mu_B$ for $^{6}$Li~\cite{liresonance}.
With a typical trapping frequency $\omega_z = 2\pi \times 62$ kHz~\cite{ol},
the dimensionless physical parameters are then given by
$g_p = 23$ $(272)$, $U_p = 1.7$ $(-5.5)$ for $^{40}$K ($^6$Li).
We notice that in the presence of SOC, the open and closed channels involved in the Feshbach resonance will be
altered since the spin is no longer a good quantum number. This modification can quantitatively change the
resonance position as well as the scattering parameters. However, as long as we express the physical
quantities in terms of the 3D scattering length $a_s$, the results would remain valid due to the requirement
of universality.

We first consider two-body bound states with zero transverse CoM momentum $q = 0$.
We show in Fig.~\ref{fig:Eb} the eigenstate energy $E_{q=0}$
of two-body bound states for (a) $^{40}$K and (b) $^{6}$Li.
For both cases, a bound state is always present around the resonance point.
This result is a combined effect of axial confinement and Rashba SOC, which both
drive the system to an effective 2D model in the low-energy limit.
For $^{40}$K, the bound state energy saturates to a limiting value on the deep BEC side,
as a direct consequence of its positive background scattering length~\cite{jason-pra06,bcsbecwy}.
The bound state energy can also be illustrated as a function of the 3D scattering length,
as shown in Fig.~\ref{fig:Eb}(c). In this plot, results for $^{40}$K and $^{6}$Li are not
distinguishable around the resonance point, indicating universal behavior at unitarity.

In order to discuss the SOC effect, we show in Fig.~\ref{fig:Eb}(d) the
binding energy of the bound state as the difference between the eigenstate energy and
the open channel threshold $-|E_b(q=0)| = E_{q=0} - E_{\rm th}$. In the presence of SOC, the open
channel consists of two colliding particles residing on the ground state of the lower
helicity band. Thus, the threshold $E_{\rm th} = -\lambda^2 +1/2$, where $1/2$ denotes the zero-point
energy of the relative degree of freedom along the confined direction.
From Fig.~\ref{fig:Eb}(d), we notice that SOC tends to enhance the two-body
binding energy throughout the entire BCS-BEC crossover region. This is understandable
considering the fact that SOC increases the single particle density of states
in the low-energy limit.

\begin{figure}[tbp]
\centering
\includegraphics[width=4.5cm]{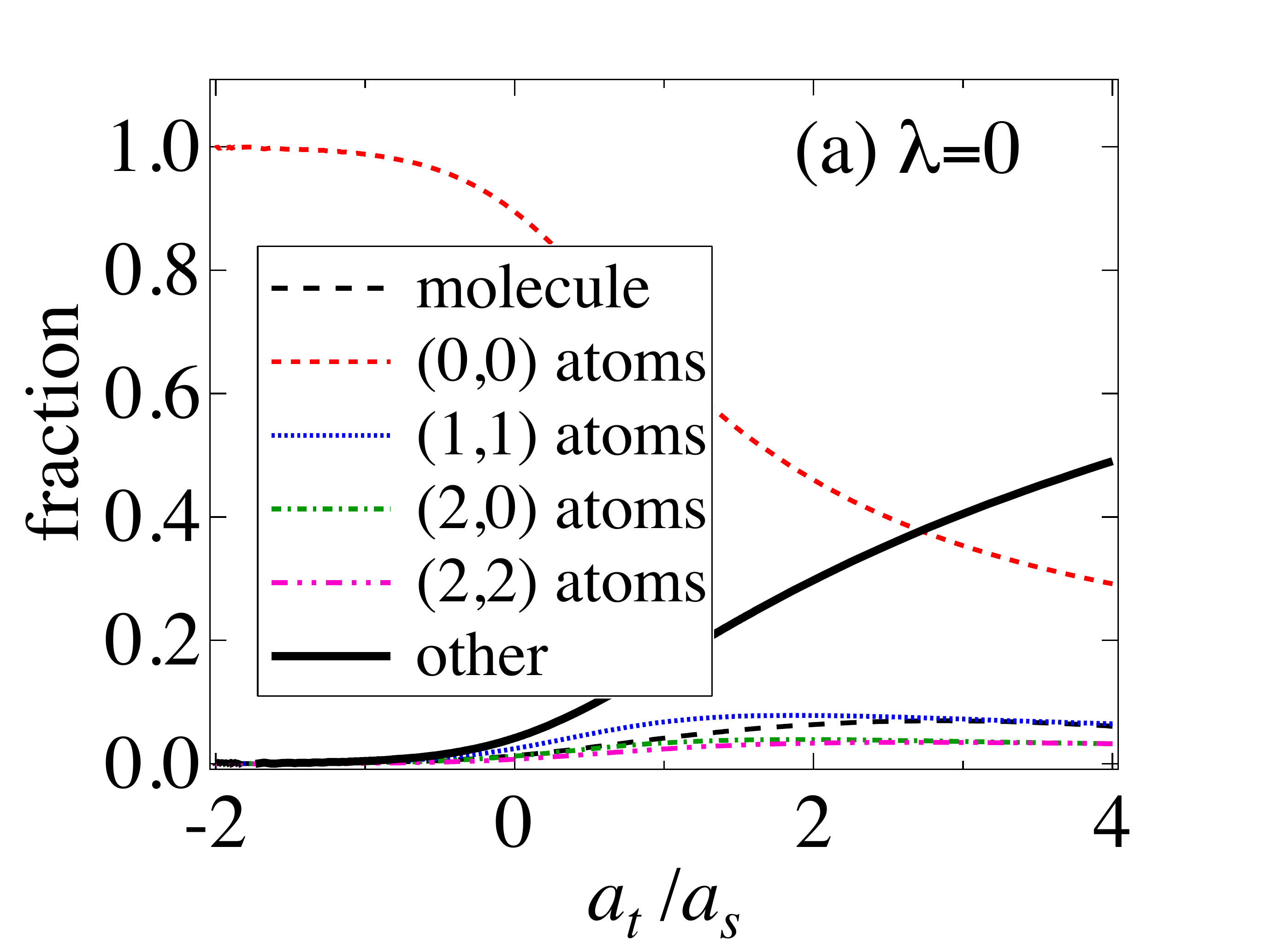}
\hskip-0.5cm
\includegraphics[width=4.5cm]{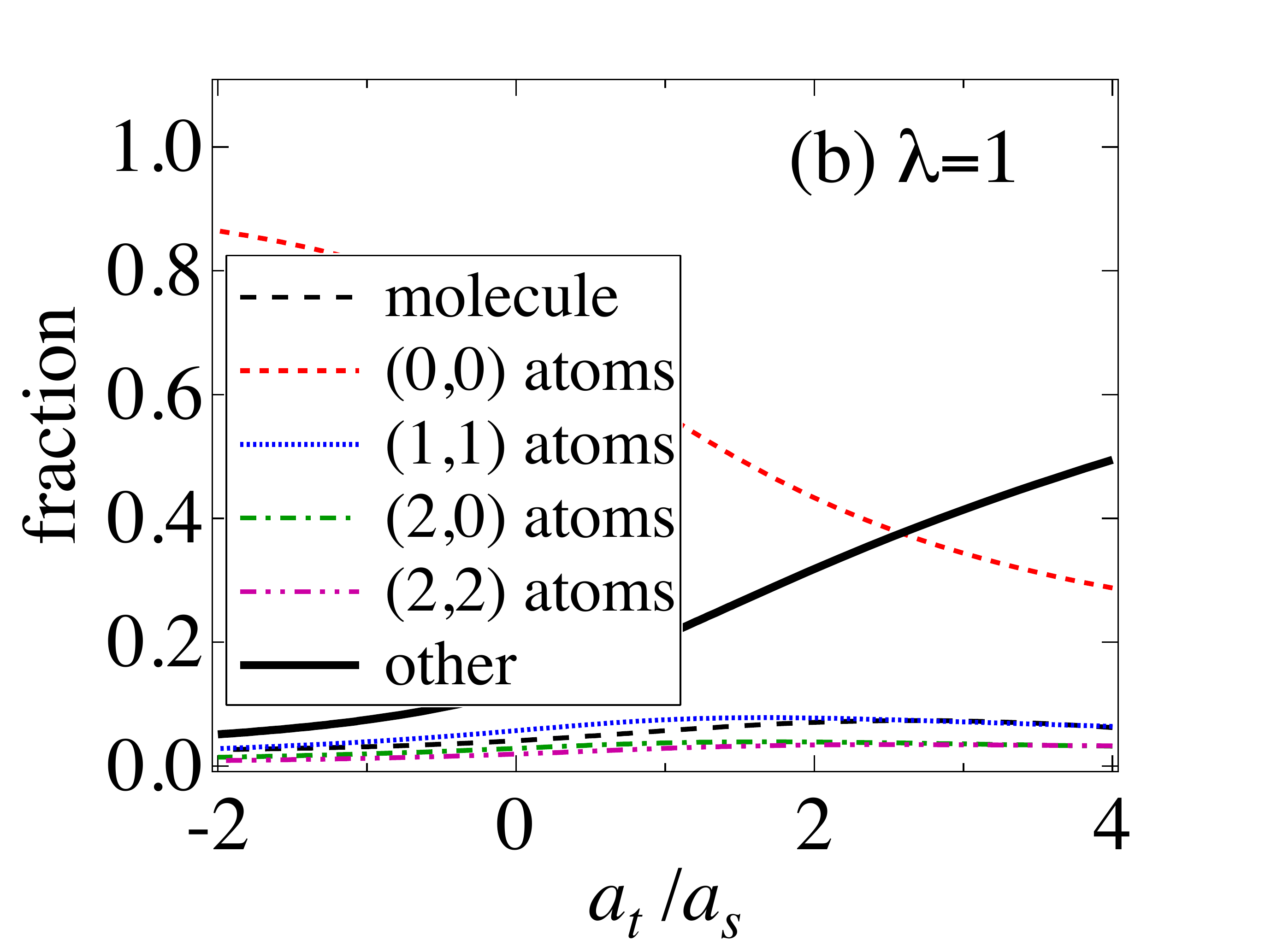}
\includegraphics[width=4.5cm]{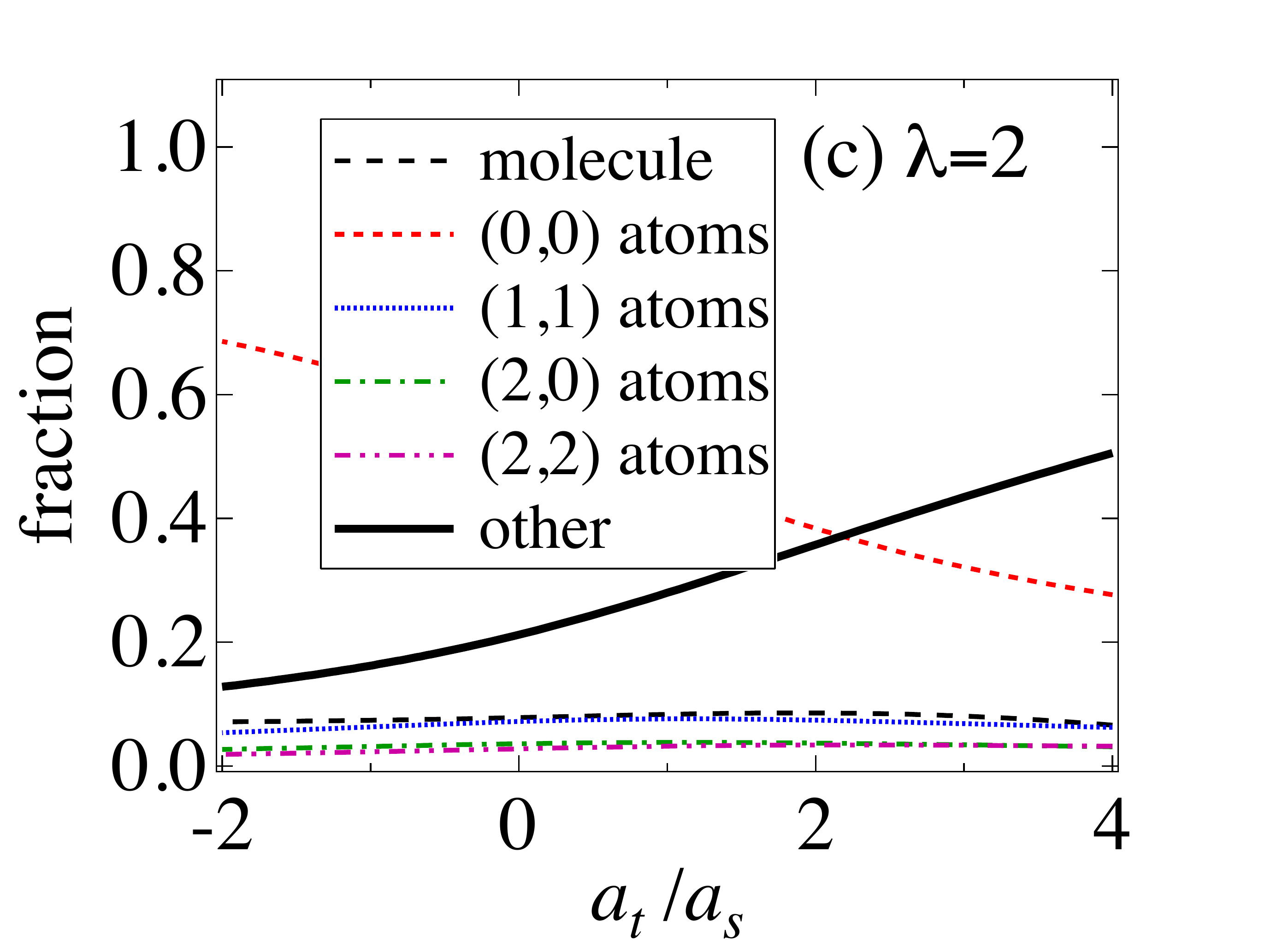}
\hskip-0.5cm
\includegraphics[width=4.5cm]{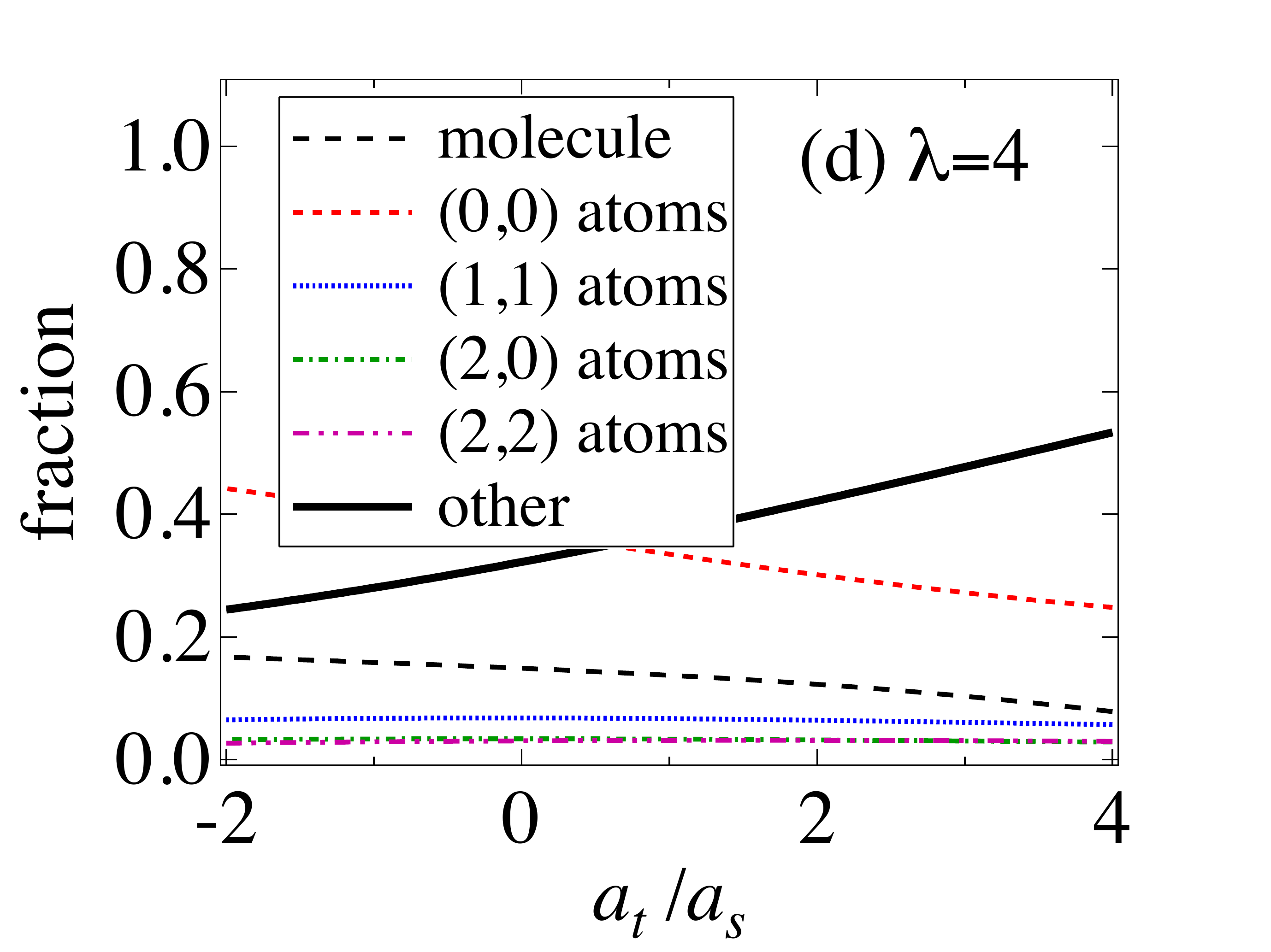}
\caption{(Color online) Population fraction of the two-body bound state in different axial harmonic levels for $^{40}$K.
Excited axial states are significantly populated around unitarity and on the BEC side
of the resonance, as a direct consequence of the increasing two-body binding energy.
The presence of SOC tends to enhance the binding energy, hence makes the population in excited states
more prominent and may become notable even on the BCS side of the resonance.
Results for $^{6}$Li are similar as plotted as functions of $a_t/a_s$.
}
\label{fig:fraction}
\end{figure}

In Fig. \ref{fig:fraction}, we show the molecular fraction $\vert \beta \vert^2$
and the population distribution in the axial harmonic levels within the two-body bound state for $^{40}$K.
The population distribution in the axial $(m,n)$ mode is a combination of all four spin states
${P}_{m,n} = \sum_{\sigma \sigma^\prime} \sum_{\bf k}^\prime
| \eta_{m,n,{\bf k},q=0}^{\sigma \sigma^\prime} |^2$.
From these results, one can see clearly that many axial excited states are occupied.
The population fraction still goes down as the energy of the modes goes up,
but the convergence is poor and there are so many excited harmonics that the
total population in the excited levels becomes significant at unitarity and eventually
dominates on the BEC side of the resonance. This result is a direct consequence of
the two-body binding energy approaching or exceeding the axial confinement, which can populate fermions to axial excited states. As SOC tends to
enhance the two-body binding energy, this effect will be more eminent with stronger SOC,
such that the population in excited harmonics can become dominant even on the BCS of the resonance.
The results for $^{6}$Li show similar behavior when plotted as functions of $a_t / a_s$.

\begin{figure}[tbp]
\centering
\includegraphics[width=4.5cm]{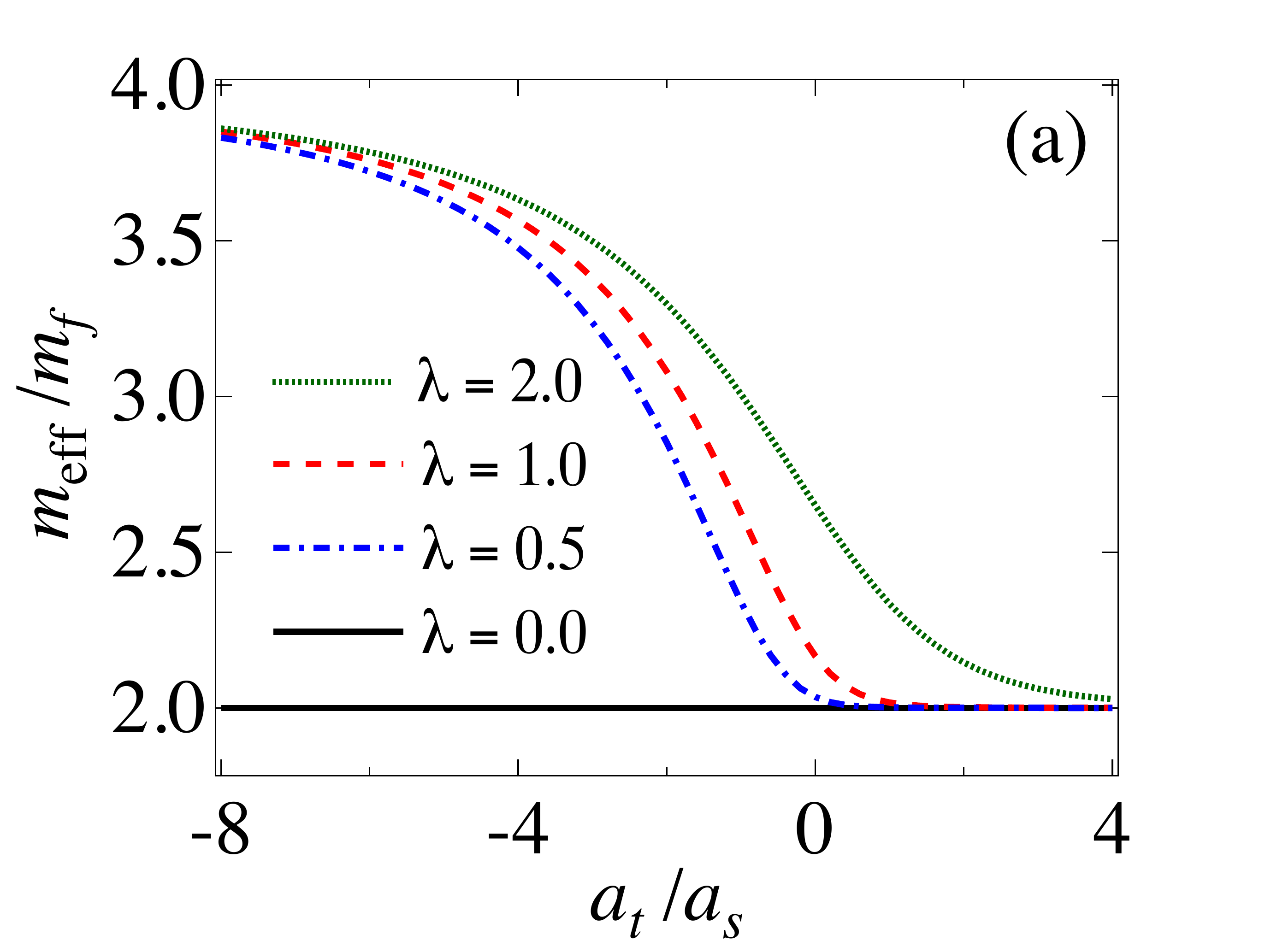}
\hskip-0.5cm
\includegraphics[width=4.5cm]{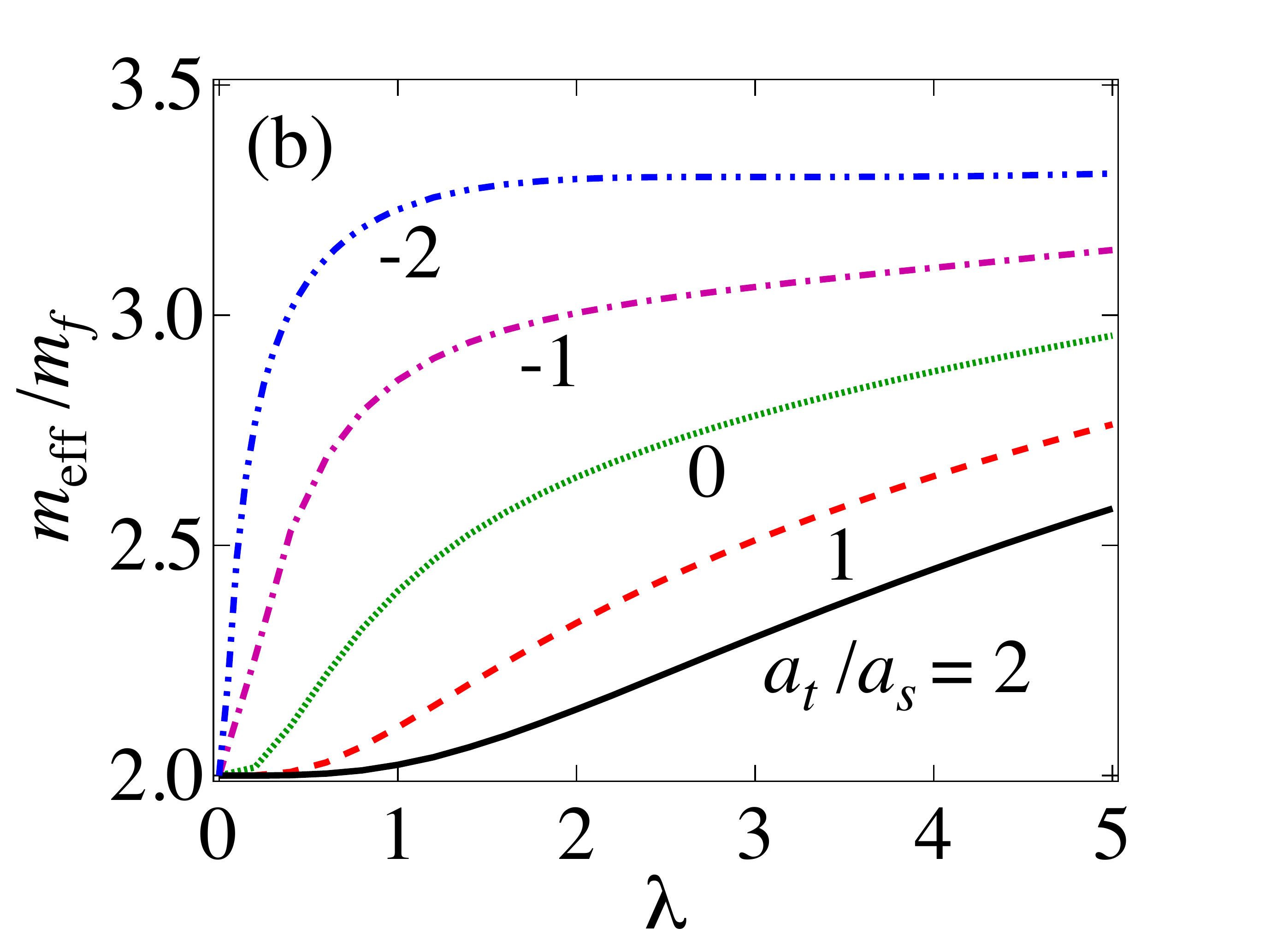}
\caption{(Color online) Effective mass $m_{\rm eff}$ of the two-body bound state for $^{40}$K.
(a) In the presence of SOC, $m_{\rm eff}$ approaches $4m_f$ in the BCS limit,
and decreases monotonically to the limiting value of $2m_f$ in the BEC limit.
(b) The effective mass becomes larger with stronger SOC intensity.
Results for $^{6}$Li are similar as functions of $a_t/a_s$.}
\label{fig:meff}
\end{figure}

Next, we discuss general two-body bound states with finite transverse CoM
momentum. For cases with small ${\bf q}$, we can expand the bound
state energy as $E_{\bf q} \approx E_{q=0} + q^2m_f /(2m_{\rm eff})$ with $m_{\rm eff}$
the effective mass of dimers. In the BCS limit with $|E_{q=0}| \ll \lambda^2$, the
molecular effective mass is $m_{\rm eff} \to 4m_f$, while in the BEC limit with
$|E_{q=0}| \gg \lambda^2$, $m_{\rm eff}$ tends to the limiting value of $2m_f$.
By tuning through the Feshbach resonance from the BCS to the BEC limit, the effective
mass decreases monotonically as shown in Fig.~\ref{fig:meff}(a). For a given
scattering length, $m_{\rm eff}$ is larger for stronger SOC, as indicated in
Fig.~\ref{fig:meff}(b). The effective mass also acquires universality
where the results for $^{40}$K and $^{6}$Li are close around
the resonance point.

Up to now, we have discussed two-body bound state in a two-component Fermi gas
with Rashba SOC confined in quasi two dimensions. We find that the axial excited
harmonic levels will be significantly populated as the dimer's binding energy
becomes comparable to or exceeds the axial confinement energy.
Compared to the case without SOC, this effect is more dramatic in the presence of SOC
as the two-body binding energy is increased. As a direct consequence, the population
of axial excited modes can be important even in the BCS regime provided that the SOC strength
is large enough. In this case, these higher excited states must be taken into account for a correct description
of the underlying system. For this purpose, next we present a 2D effective model which
incorporates the effect of axial exited states into a so-called {\it dressed molecule} state.
This model can describe the quasi-2D, but truly 3D Fermi gas in the low-energy limit.
%

%%%%%%%%%
\section{Dressed molecule and effective 2D Hamiltonian}
\label{sec:2DH}

In order to write down the correct effective 2D theory, we notice that the original quasi-2D
Hamiltonian Eq. (\ref{eqn:H1}) contains three types of degree-of-freedom (DoF), including fermions
in the axial ground state, fermions in axial excited states, and Feshbach molecules.
These three types of DoF can be categorized into different length and energy scales. For fermions
in the axial ground state, the corresponding length and energy scales are the inter-particle separation $d$
and the Fermi energy $E_F$, respectively. For fermions in axial excited states, the corresponding
length and energy scales are $a_t$ and $\hbar \omega_z$. The Feshbach molecular state
is related to the short-range details of the atom-atom interaction potential, which has length
scale $R_e$ and energy scale $V_e$. In ultra-cold Fermi gases with quasi-2D confinement,
these three length and energy scales are usually well separated, satisfying $d \gg a_t \gg R_e$
and $E_F \ll \hbar\omega_z \ll V_e$.
This observation allows us to write down an effective Hamiltonian to describe physics in the
long-wavelength and low-energy limit.

We combine the high-energy DoF including fermions in axial excited states
and the Feshbach molecule to define a molecular state, which we call the {\it dressed} molecule
since it can be viewed as the Feshbach molecule dressed by excited fermions~\cite{jason-pra06, Duan-07}.
Notice that the dressed molecule is structureless since all short-range details associated with
excited fermions and the Feshbach molecule become irrelevant in the low-energy regime.
With this dressed molecule and the fermions in the axial ground state, we write down an effective
2D Hamiltonian in the form of a two-channel model (also in dimensionless form with
length unit $a_t$ and energy unit $\hbar \omega_z$)
\begin{eqnarray}
\label{eff-H}
H_{\rm eff} &=&
\sum_{{\bf k},\sigma} \epsilon_{\bf k} a_{{\bf k},\sigma}^\dagger a_{{\bf k}, \sigma}
+ \sum_{\bf q} \left(\delta_b + \epsilon_{\bf q}/2 \right) d_{\bf q}^\dagger d_{\bf q}\nonumber\\
&&\hspace{-1cm}
+ \lambda^\prime \sum_{\bf k} \left[
(k_x - ik_y) a_{{\bf k}, \uparrow}^\dagger  a_{{\bf k}, \downarrow}
+ {\rm H.C.} \right]
\nonumber \\
&&\hspace{-1cm}
+ \alpha_b \sum_{{\bf k},{\bf q}} \left( d_{\bf q}^\dagger a_{{\bf k}+{\bf q}/2,\uparrow}
a_{-{\bf k}+{\bf q}/2,\downarrow}  + {\rm H.C.} \right)\nonumber\\
&&\hspace{-1cm}
+ V_b \sum_{{\bf k},{\bf k}^\prime,{\bf q}}
a_{{\bf k}+{\bf q}/2,\uparrow}^\dagger a_{-{\bf k}+{\bf q}/2,\downarrow}^\dagger
a_{-{\bf k}^\prime+{\bf q}/2, \downarrow} a_{{\bf k}^\prime+{\bf q}/2, \uparrow}.
\end{eqnarray}
Here, $a_{{\bf k},\sigma}^\dagger (a_{{\bf k}, \sigma})$ are fermionic creation (annihilation) operator
in two dimensions, $\epsilon_{\bf k} = k^2/2$ is the corresponding dispersion relation,
$d_0^\dagger$ and $d_0$ denote the dressed molecular operators, and the Rashba SOC
is characterized by the effective coupling constant  $\lambda^\prime$.
The three bare scattering parameters $\delta_b$, $\alpha_b$, and $V_b$ can be linked to
physical ones via a 2D renormalization analogous to Eq. (\ref{eqn:renorm})
\begin{eqnarray}
\label{eqn:renorm2}
&&V_{c}^{-1} = - \int \frac{d^2{\bf k}}{(4\pi^2)} \frac{1}{2 \epsilon_{\bf k}+1},
\quad
\Omega^{-1} = 1 + \frac{{V}_p}{V_c},
\nonumber \\
&&{V}_p = \Omega^{-1} {V}_b, \quad {\alpha}_p = \Omega^{-1} {\alpha}_b, \quad
{\delta}_p = {\delta}_b + \Omega \frac{{\alpha}_p^2}{V_c}.
\end{eqnarray}

The parameters in the effective Hamiltonian are determined by the following considerations.
First, we require the effective model to give the correct open channel threshold, which represents
the energy of two particles when they are far apart. In other words, the effective theory needs to
reproduce the single-particle dispersion in the low-energy limit. Second, when the system is far away from
resonance where the population of both the Feshbach molecule and excited fermions are negligible,
the quasi-2D Fermi gas can be well described by a 2D system with all fermions staying in the axial ground
state. This observation indicates that the background interaction $V_b$ in the effective 2D theory should be
related to the 3D background interaction $U_b$ by integrating out the harmonic ground state
wavefunction along the axial direction.

By tuning through the Feshbach resonance, we further require the effective 2D model to
reproduce the correct two-body physics. In fact, since the possibility for three and more particles come
to a close range is exponentially small in dilute atomic gases, the low-energy physics
is dominated by two-body processes. For the quasi-2D system around the BCS-BEC crossover,
the ground state is always a two-body bound state with axial CoM mode $\ell = 0$ and
transverse CoM momentum $q=0$. Therefore,
the effective 2D theory should give the same two-body binding energy $|E_b(\ell = 0, q =0)|$
as the original Hamiltonian. Besides, since the dressed molecule is a phenomenological combination
of the Feshbach molecule and axial excited fermions, we also require the population of dressed molecule
to match the population of Feshbach molecule plus that of the dimers formed with axial excited fermions.
\begin{figure}[tbp]
\centering
\includegraphics[width=4.5cm]{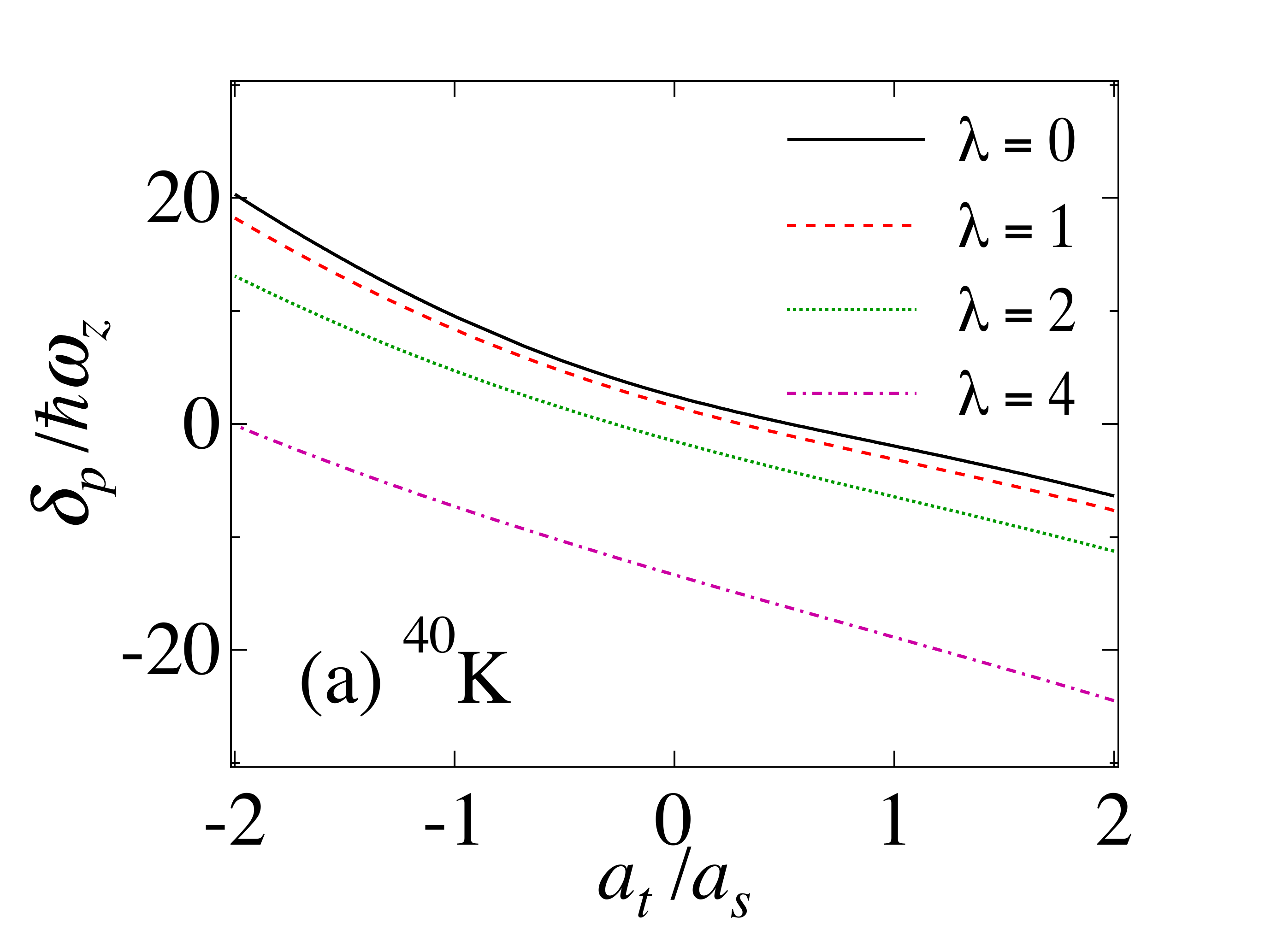}
\hskip-0.5cm
\includegraphics[width=4.5cm]{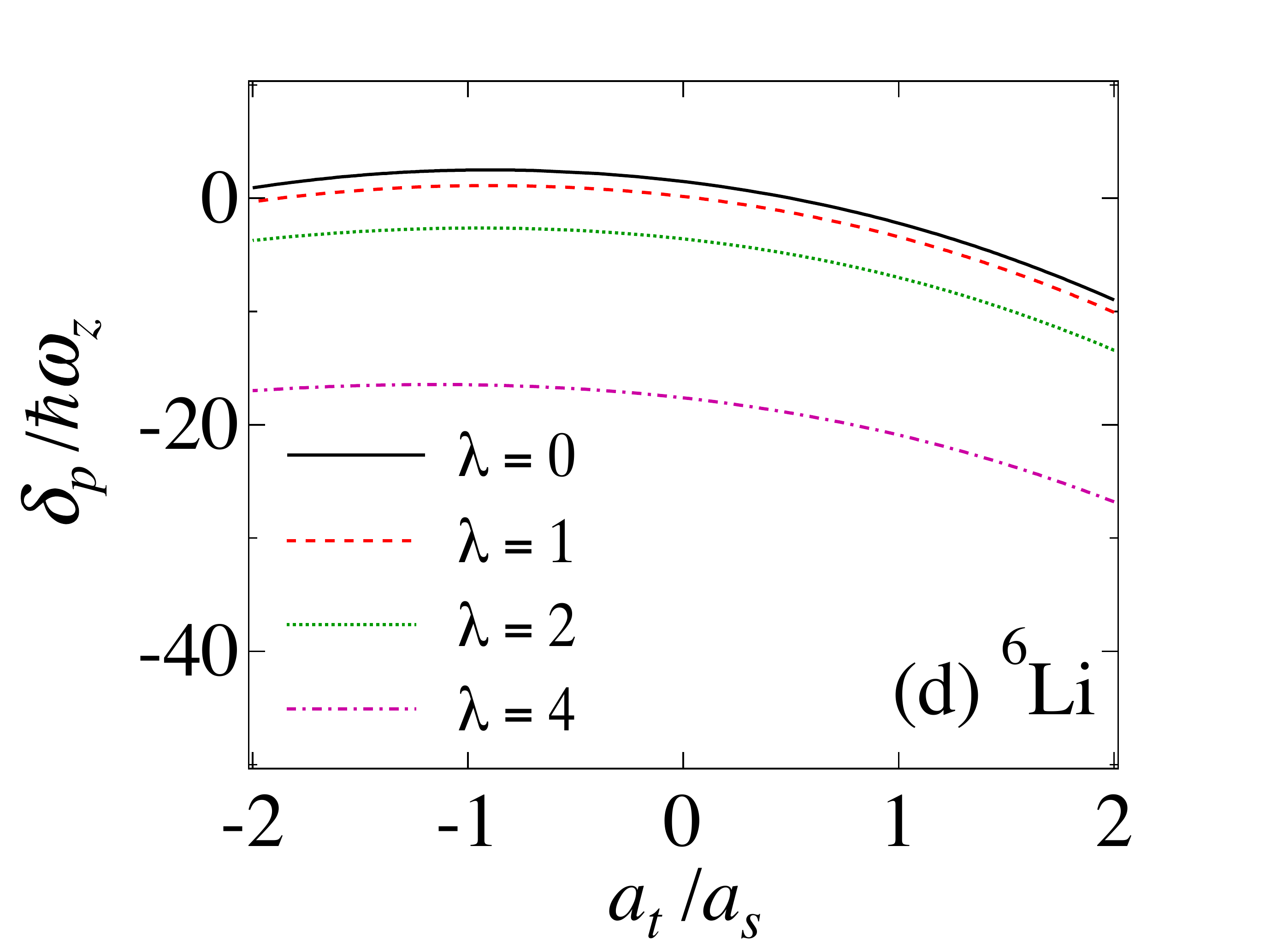}
\includegraphics[width=4.5cm]{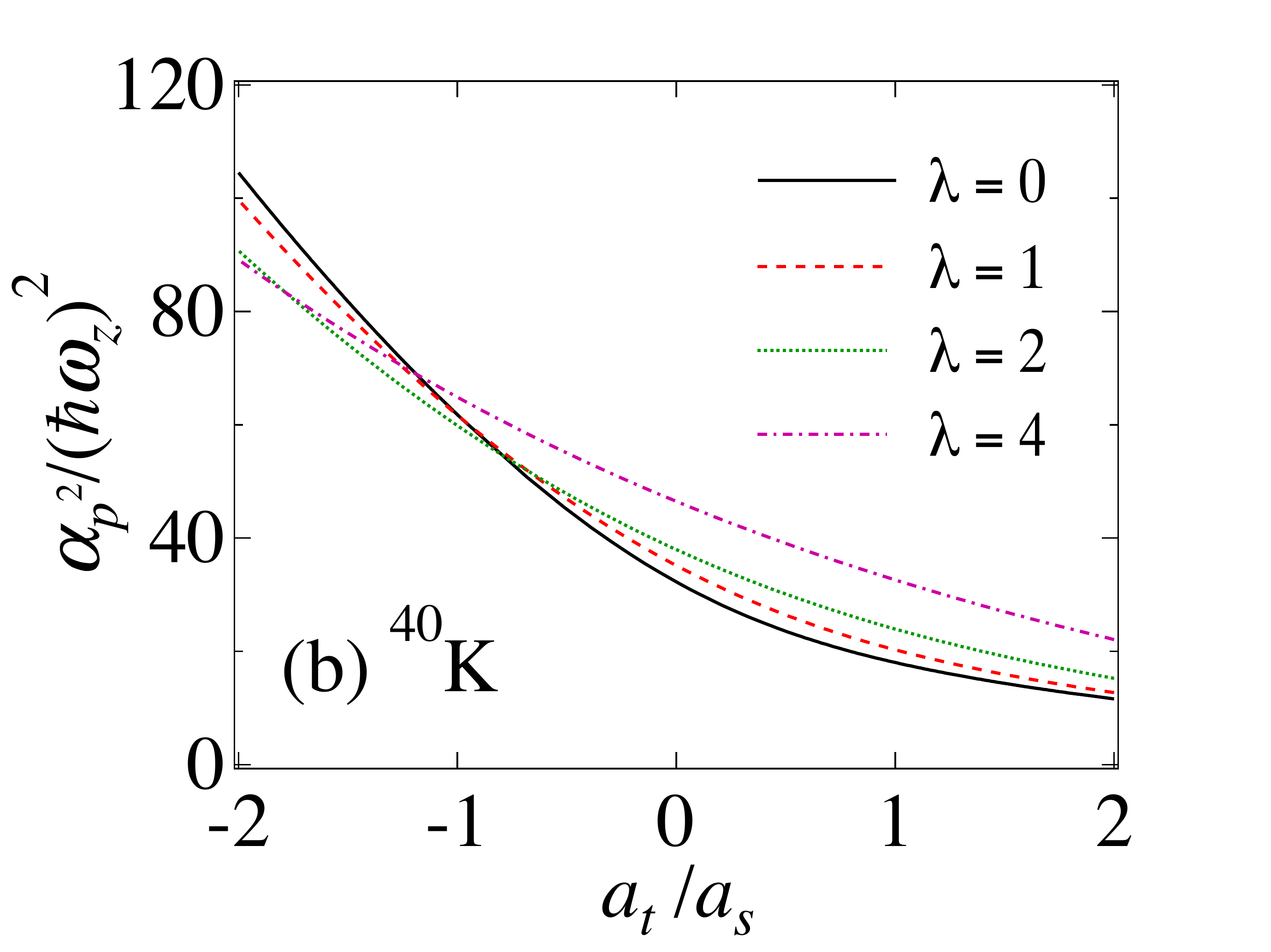}
\hskip-0.5cm
\includegraphics[width=4.5cm]{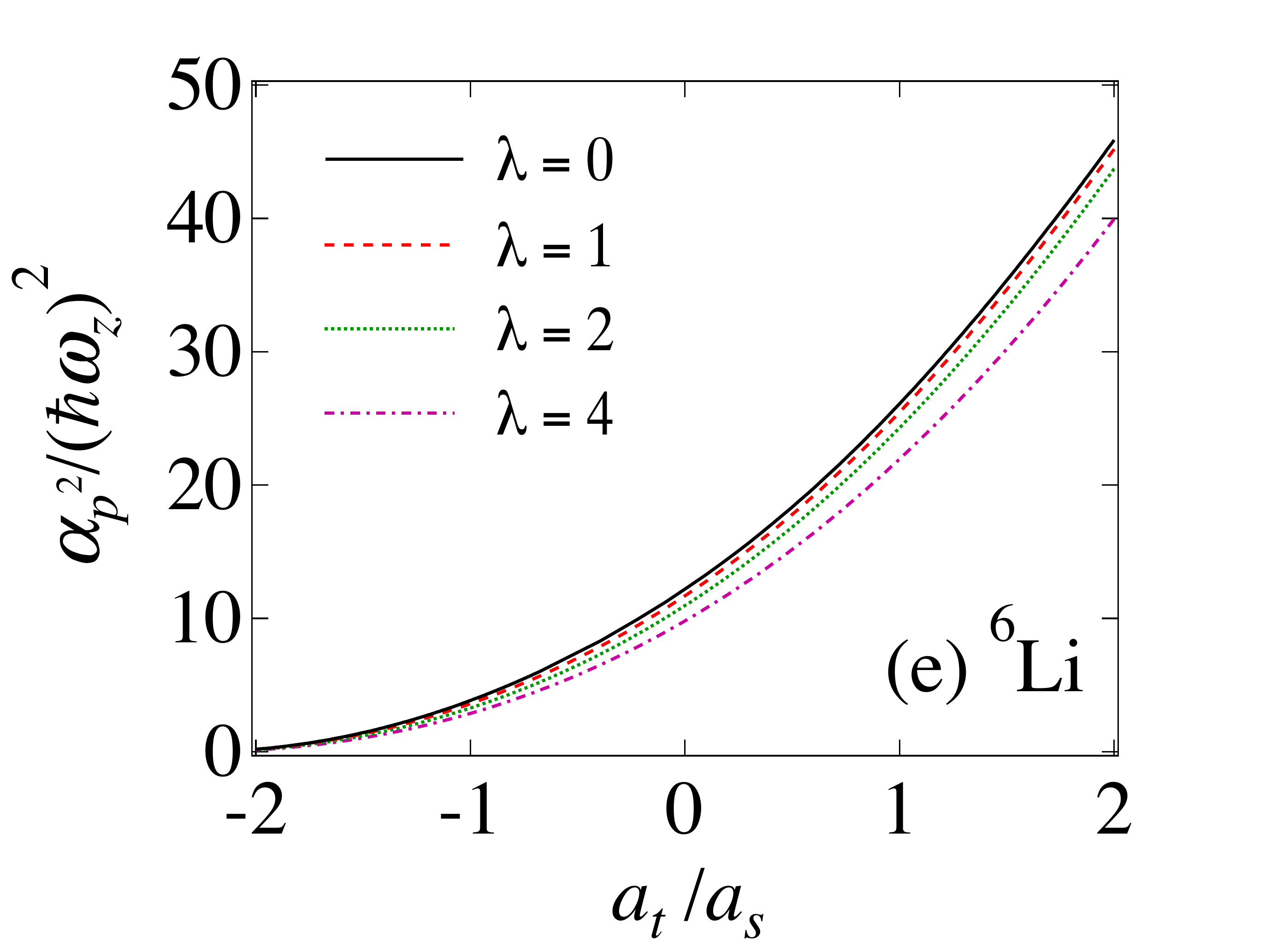}
\includegraphics[width=4.5cm]{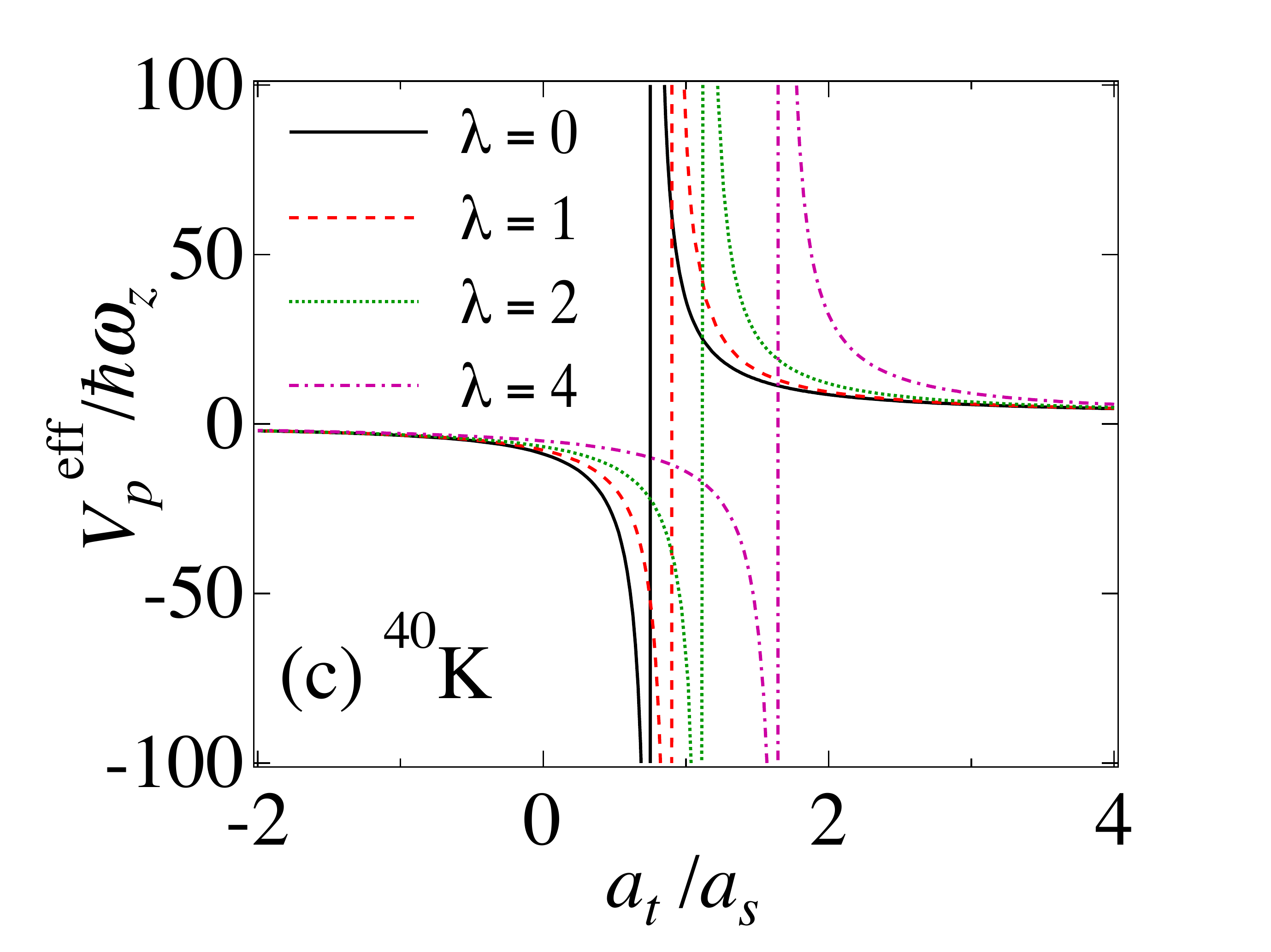}
\hskip-0.5cm
\includegraphics[width=4.5cm]{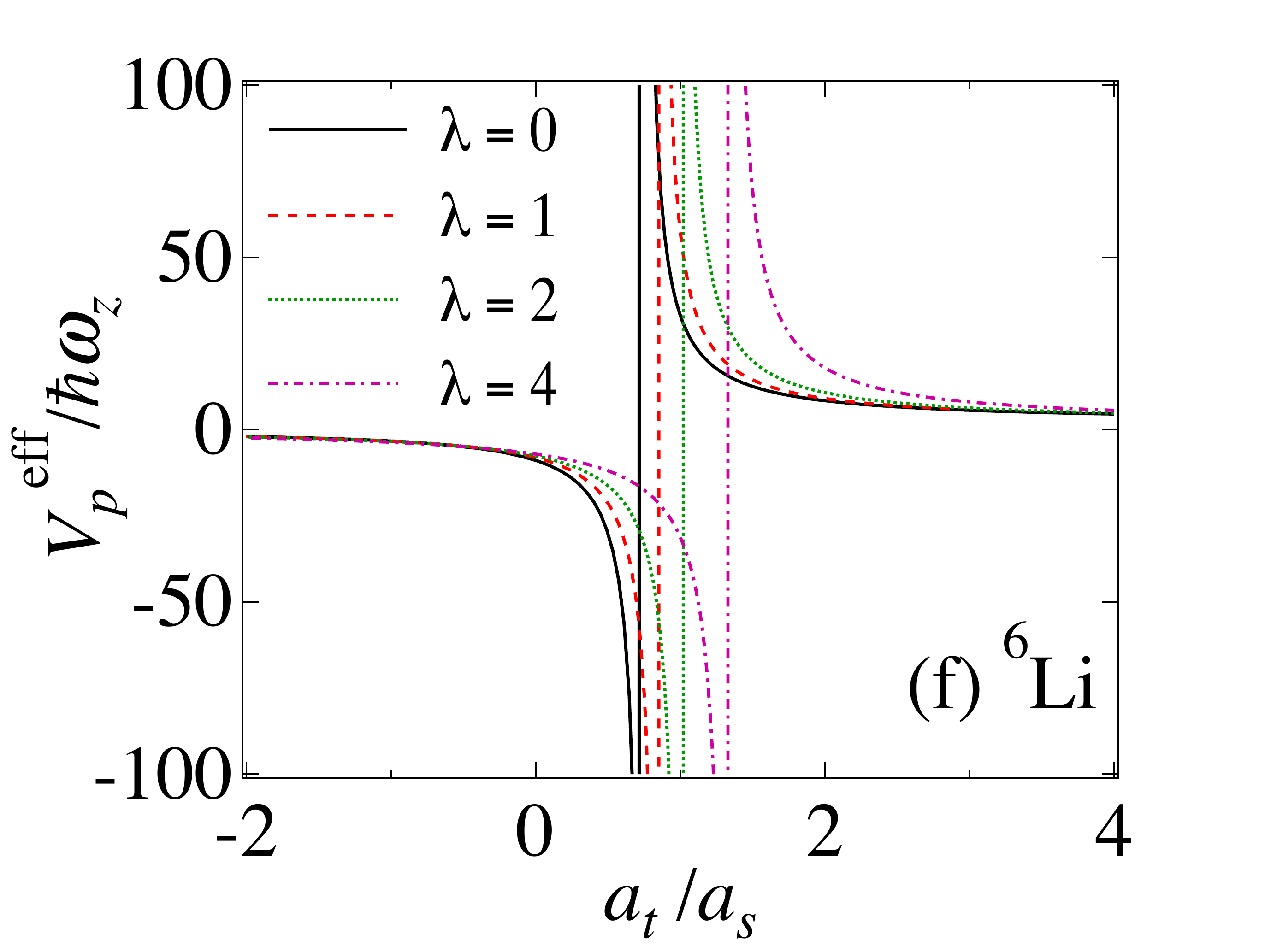}
\caption{(Color online) Parameters in the effective 2D Hamiltonian as functions of $a_t/a_s$.
The energy independent background scattering
amplitude $V_p$ for different SOC constants are: 3.134 ($\lambda$=0), 3.147 ($\lambda$=1), 3.186 ($\lambda$=2), 3.341 ($\lambda$=4) for $^{40}$K, and -1.801 ($\lambda$=0), -1.828 ($\lambda$=1), -1.914 ($\lambda$=2),
-2.155 ($\lambda$=4) for $^{6}$Li.}
\label{fig:para}
\end{figure}

To study the two-body problem within the effective 2D model, we take the ansatz wavefunction as
\begin{eqnarray}
\label{eqn:2Dphi}
| \Phi \rangle_{\bf q} &=& \left( \beta d_{\bf q}^\dagger+ \sum_{\bf k} \ ^\prime \sum_{\sigma,\sigma^\prime}
\eta_{\bf k}^{\sigma \sigma^\prime} a_{{\bf k}+{\bf q}/2,\sigma}^\dagger
a_{-{\bf k}+{\bf q}/2,\sigma^\prime}^\dagger
\right)| 0 \rangle,\nonumber\\
\end{eqnarray}
where $\sum_{\bf k}^\prime$ indicates summation over 2D momentum with $k_y>0$,
and the summation over spins runs over all four combinations of $(\sigma, \sigma^\prime)$.
We focus on the case with $q=0$ and substitute the wavefunction into the Schr{\"o}dinger's equation
$H_{\rm eff} | \Phi \rangle_{q=0} = E_b | \Phi \rangle_{q=0}$, which leads to
\begin{eqnarray}
\left[ V_p - \frac{\alpha_p^2}{ \delta_p - E_b} \right]^{-1} = \sqrt{2 \pi } \sigma_p(E_b),
\end{eqnarray}
where the function $\sigma_p$ is defined as
\begin{eqnarray}
\sigma_{p}(E_b) &=& \int \frac{d^{2}{\bf k}}{(2\pi)^{5/2}}
\left[ \frac{1}{E_b - 2 \epsilon_{\bf k} - \frac{4 (\lambda^\prime)^2 k^2 }{E_b-2 \epsilon_{\bf k}}}
+ \frac{1}{1+ 2 \epsilon_{\bf k}} \right]
\nonumber \\
&& \hspace{-1cm}
- b \frac{\pi + 2 {\rm tan}^{-1}\left( \frac{b + 2E_b}{\sqrt{-b (4E_b + b)}} \right)}{8 \pi \sqrt{2 \pi} \sqrt{-b(4E_b + b)}}
+ \frac{\ln(-E_b)}{4 \pi \sqrt{2 \pi}}
\end{eqnarray}
and $b \equiv 4 (\lambda^\prime)^2$ is used to simplify notation.

With these results, we can fix the parameters in the effective 2D Hamiltonian by matching
single- and two-body physics as discussed above
\begin{eqnarray}
\lambda^\prime &=& \lambda,
\label{eqn:para1}
\nonumber \\
V_{p}^{-1} &=& \sqrt{2 \pi} \left(U_p^{-1} - C_p \right),
\label{eqn:para2}
\nonumber \\
\delta_p &=& E_b - \frac{\sigma_p(E_b)}{\partial {\cal P}(E_b) /\partial E_b}
\left( 1- \frac{\sigma_p(E_b)}{ U_p^{-1} - C_p} \right),
\label{eqn:para3}
\nonumber \\
\alpha_p^2 &=& \frac{1}{ \sqrt{2 \pi} {\partial {\cal P}(E_b) /\partial E_b}}
\left( 1- \frac{\sigma_p(E_b)}{ U_p^{-1} - C_p} \right)^2.
\label{eqn:para4}
\end{eqnarray}
Here, the parameters are defined as
\begin{eqnarray}
C_p &=& S_p(E_b^{\rm inf}+1/2) - \sigma_p(E_b^{\rm inf}),
\nonumber \\
{\cal P} (E_b) &=& \left[ 1/U_p^{\rm eff} - S_p(E_b+1/2) + \sigma_p(E_b)  \right],
\nonumber \\
U_p^{\rm eff} &=& U_p - g_p^2/(\nu_p - E_b),
\end{eqnarray}
with $E_{b}^{\rm inf}$ the two-body binding energy in quasi two dimensions for $\nu_p \to \infty$.
Notice that the eigenenergy of the two-body bound state as determined by
Eq. (\ref{eqn:bindEeqn}) is shifted from the binding energy by the zero-point energy
along the strongly confined axial direction.

Equations (\ref{eqn:para4}), together with the renormalization relation Eq. (\ref{eqn:renorm2}),
fix the parameters in the effective 2D Hamiltonian (\ref{eff-H}) as functions of two-body binding
energy $E_b$, which is tuned via the physical detuning $\nu_p$ through the Feshbach resonance.
In Fig.~\ref{fig:para}, we plot the parameters $\delta_p$ and $\alpha_p$ for $^{40}$K and $^{6}$Li
across resonance for various SOC strengths, using the same 3D parameters as in Fig.~\ref{fig:Eb}.
These effective parameters behave qualitatively differently for $^{40}$K and $^{6}$Li, mainly
due to the difference in sign of their individual background interaction. We also show
the effective interaction $V_p^{\rm eff} \equiv V_p - \alpha_p^2/(\delta_p - E_b)$, as it will be used
for the mean-field calculation discussed below.

Up to now, we have introduced an effective 2D Hamiltonian by matching single-
and two-body physics with the original quasi-2D system. This effective theory is derived
by grouping the high-energy DoF of axial excited fermions and Feshbach molecules
to define a dressed molecular state, while keeping the low-energy axial ground fermions
to catch the correct low-energy physics of the corresponding many-body system.
To justify the validity of this effective theory, we notice that in the language of $T$-matrix,
the matching conditions of two-body binding energy and molecular fraction correspond
to matching the pole of two-body $T$-matrix and the first derivative of $T^{-1}$ around the pole, respectively.
Thus, we conclude that as the fermionic chemical potential $2\mu$ is not far away from the
bound state energy $- |E_b|$, the $T$-matrix of the original quasi-2D Hamiltonian can
be well approximated by that of the effective 2D model
\begin{eqnarray}
T(x) \approx T_{\rm eff} (x),
\quad {\rm as\ } |x| \equiv \left| 2\mu + |E_b| \right| \ll 1.
\end{eqnarray}
Considering the fact that $2\mu + |E_b|$ is of the same order as the Fermi energy $E_F$ through
the BCS-BEC crossover, and that the Fermi energy is proportional to the 2D number density
$E_F \propto n_{\rm 2D}$, we conclude that the effective theory is approximately valid as the
diluteness and the quasi-2D conditions hold for the underlying Fermi gas.

%%%%%%%%%

\section{Many-body calculations with the effective two-channel model}

With the bare parameters of the effective two-channel model fixed, we may now proceed with the many-body calculations and demonstrate the significance of the dress molecules. Here, we consider only the possibility of BCS superfluid and focus on the zero CoM momentum pairing state with $q=0$.
The dimensionless many-body Hamiltonian can be written as
\begin{eqnarray}
&&H_{\rm eff}-\mu \left(N_{\uparrow}+N_{\downarrow}\right)-h\left(N_{\uparrow}-N_{\downarrow}\right)
\nonumber \\
&& \hspace{5mm}
=\sum_{\mathbf{k},\sigma}\left(\epsilon_{\mathbf{k}}-\mu_{\sigma}\right)a^{\dag}_{\mathbf{k},\sigma}a_{\mathbf{k},\sigma} +(\delta_b-2\mu) d^{\dag}_0d_0
\nonumber\\
&&\hspace{5mm}
+\lambda \sum_{\mathbf{k}}\left[(k_x-i k_y)a^{\dag}_{\mathbf{k},\uparrow}a_{\mathbf{k},\downarrow}+(k_x+ik_y) a^{\dag}_{\mathbf{k},\downarrow}a_{\mathbf{k},\uparrow}\right]
\nonumber\\
&&\hspace{5mm}
+\alpha_b\sum_{\mathbf{k}}\left(d^{\dag}_0a_{\mathbf{k},\uparrow}a_{-\mathbf{k},\downarrow}+{\rm H.C.}\right) \nonumber \\
&&\hspace{5mm}
+V_b\sum_{\mathbf{k},\mathbf{k}'}a^{\dag}_{\mathbf{k},\uparrow}a^{\dag}_{-\mathbf{k},\downarrow}a_{-\mathbf{k}',\downarrow}a_{\mathbf{k}',\uparrow},
\end{eqnarray}
where $N_{\uparrow}$ ($N_{\downarrow}$) is the total number of particles in the corresponding spin state, and the bare parameters are fixed by the renormalization relation Eq. (\ref{eqn:renorm2}), together with Eqs. (\ref{eqn:para4}). Assuming a slow-varying harmonic potential in the $x$-$y$ plane, we take the local density approximation (LDA) in the transverse direction such that $\mu(\mathbf{\tilde{r}})=\mu(0)-V(\mathbf{\tilde{r}})$, where $V(\mathbf{\tilde{r}})$ is the dimensionless external trapping potential, $\mu(\mathbf{\tilde{r}})$ is the local chemical potential, $h$ is the effective Zeeman field. To be consistent with the two-body calculations, we have taken the unit of energy to be $\hbar\omega_z$, and the unit of length along the $i$th ($i=x,y$) direction to be $R_i=a_t\omega_z/\omega_i$. The dimensionless radial harmonic potential is then $V(\mathbf{\tilde{r}})=\tilde{r}^2/2$, with $\tilde{r}^2=\sum_i r_i^2/R_i^2$.
\begin{figure}[tbp]
\centering
\includegraphics[width=4.5cm]{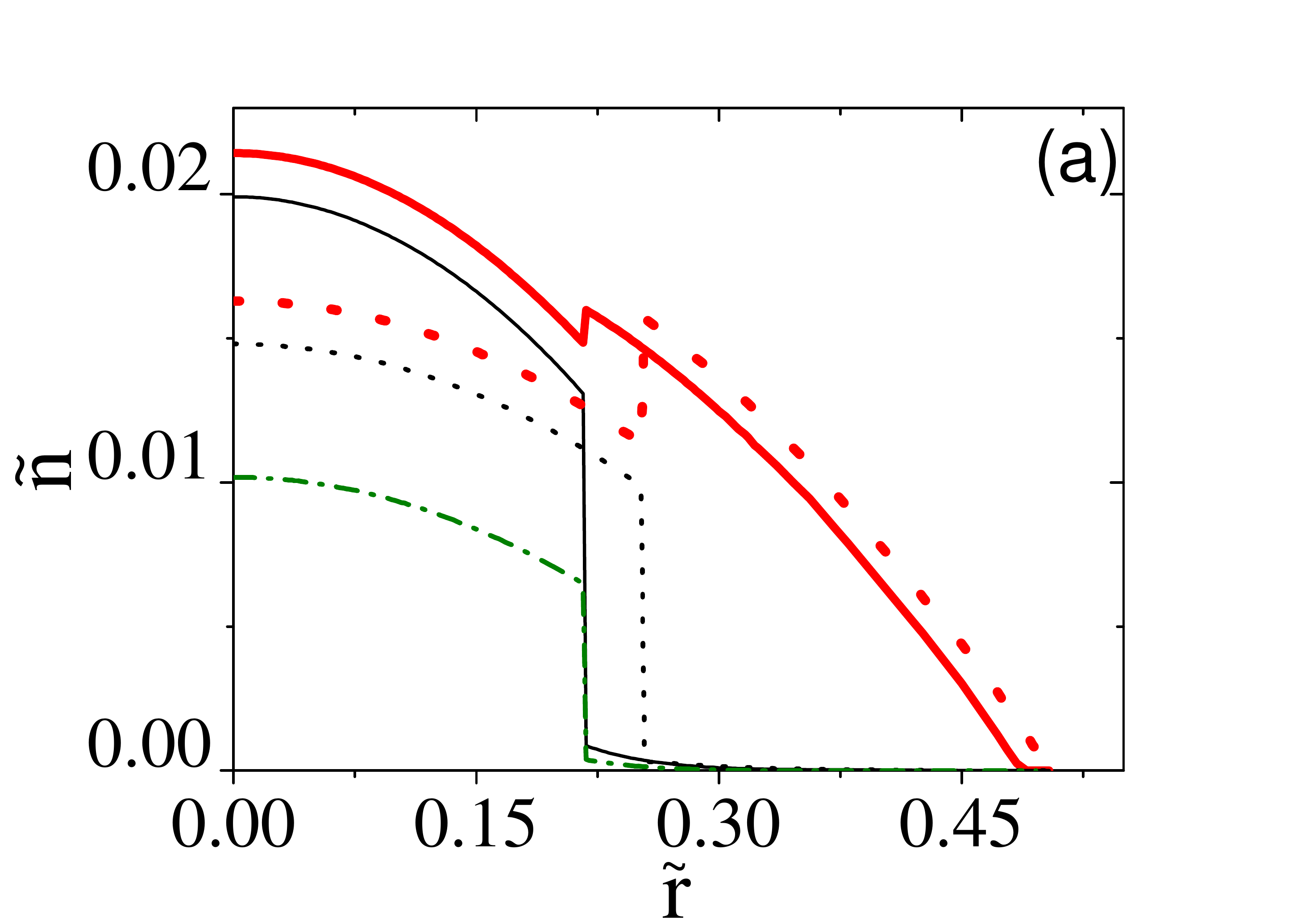}
\hskip -0.5cm
\includegraphics[width=4.5cm]{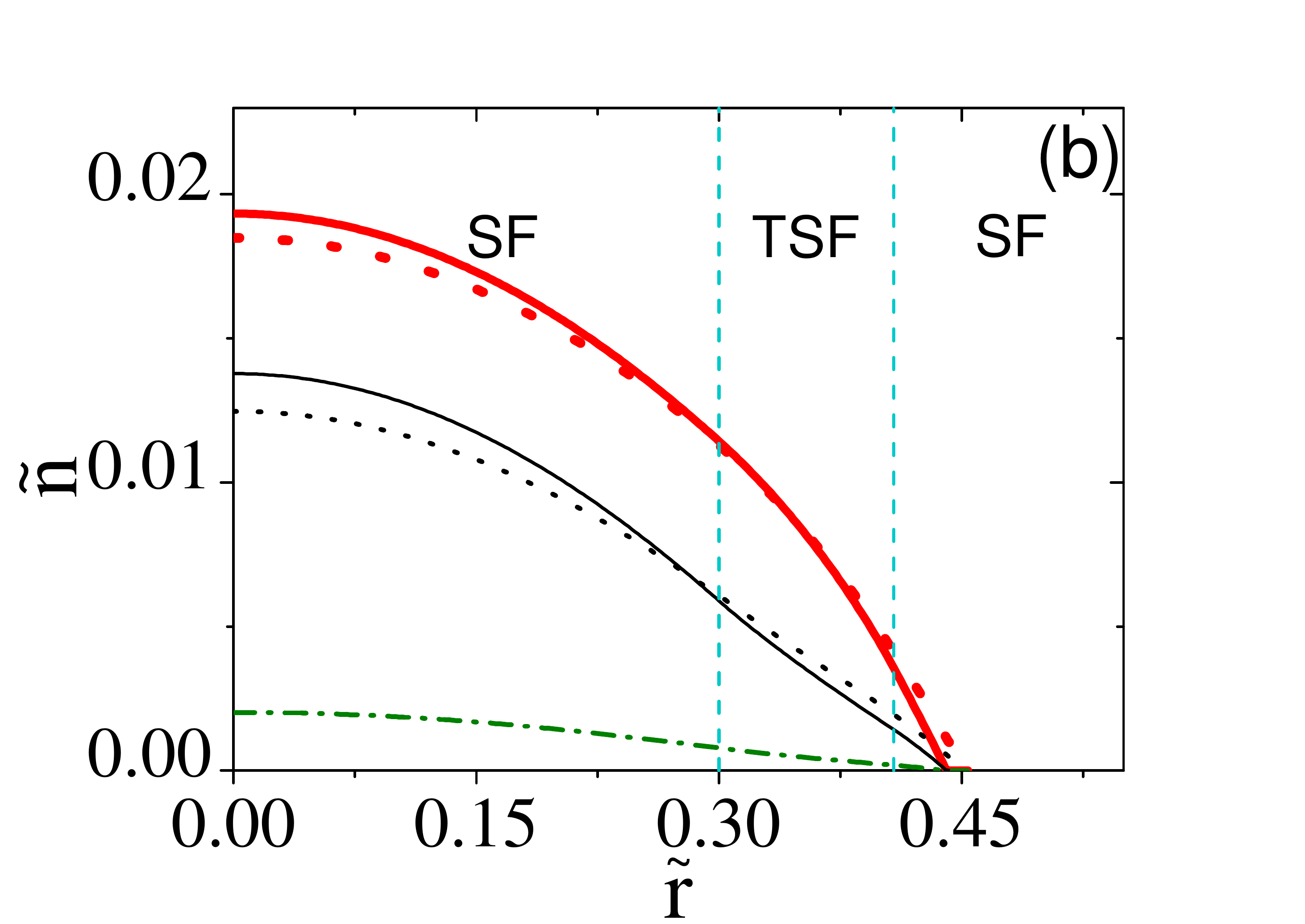}
\caption{(Color online) Density distribution for spin-up (bold, red) and spin-down (thin, black) atoms in the single-channel (dotted) and the effective two-channel (solid) models, with parameters: (a) $a_t/a_s\sim 0.34$, $\lambda\sim 0.14$, $\omega_x/\omega_z=10^{-3}$, $N\sim 10^4$, $P\sim 0.50$, $2N_b/N\sim 13\%$; (b) $a_t/a_s\sim -0.40$, $\lambda\sim 0.22$, $\omega_x/\omega_z=10^{-3}$, $N\sim 10^4$, $P\sim 0.27$, $2N_b/N\sim 5\%$. For the effective two-channel calculation, the percentage of atoms in the closed channel $2N_b/N=2\int d^2\mathbf{r} n_b(\mathbf{r})/N$, where the density distribution of the closed channel atoms $2n_b$ are shown with the dash-dotted line (green). The first-order phase boundaries in (a) between the SF and TSF phases are indicated by the abrupt jump in the density distributions. The vertical dashed lines in (b) indicate the second-order phase boundaries between the SF and TSF phases in the two-channel model.}\label{figden1}
\end{figure}

Following Ref.~\cite{jzhou}, the zero-temperature thermodynamic potential can be evaluated on the mean-field level
\begin{eqnarray}
\Omega=-\frac{|\Delta|^2}{V_p^{\rm eff}}+\frac{1}{2}
\sum_{\mathbf{k},s = \pm} \left( \xi_{s}- E_{\mathbf{k},s} \right),
\end{eqnarray}
where the order parameter for the two-channel model $\Delta=\alpha_p \langle d_0\rangle+V_p\sum_{\mathbf{k}}\left\langle a_{-\mathbf{k},\downarrow}a_{\mathbf{k},\uparrow}\right\rangle$, and $V_p^{\rm eff} \equiv V_p - \alpha_p^2/(\delta_p - E_b)$.
The dispersion relations for quasi-particles in the presence of SOC are given as \cite{jzhou}
\begin{equation}
E_{\mathbf{k},\pm}=\sqrt{\xi^2_{\mathbf{k}}+\lambda^2k^2+|\Delta|^2+h^2\pm2E_0},
\end{equation}
where $E_0=\sqrt{h^2\left(\xi_{\mathbf{k}}^2+|\Delta|^2\right) + \lambda^2 \xi_{\mathbf{k}}^2 k^2}$
and $\xi_{\bf k} = \epsilon_{\bf k} - \mu(\mathbf{\tilde{r}})$.
Due to the competition between pairing and polarization, one needs to find the global minimum of the thermodynamic potential to avoid getting meta-stable or unstable states.

The local chemical potentials can be determined self-consistently from the dimensionless number equations
\begin{eqnarray}
&&\tilde{N}=\int d^2\tilde{\mathbf{r}}[\tilde{n}_{\uparrow}(\tilde{\mathbf{r}})+\tilde{n}_{\downarrow}(\tilde{\mathbf{r}})], \label{no1}\\
&&P=\frac{1}{\tilde{N}}\int d^2\tilde{\mathbf{r}}[\tilde{n}_{\uparrow}(\tilde{\mathbf{r}})-\tilde{n}_{\downarrow}(\tilde{\mathbf{r}}))],\label{no2}
\end{eqnarray}
where $\tilde{n}_{\uparrow}=-(\partial \Omega/\partial \mu+\partial \Omega/\partial h)/2$, $\tilde{n}_{\downarrow}=-(\partial \Omega/\partial \mu-\partial \Omega/\partial h)/2$, $\tilde{N}=N\omega_x\omega_y/\omega_z^2$, with $N$ the total particle number in the trap. Solving these equations while minimizing the local thermodynamic potential, we get density distributions of the gas in a typical quasi-2D trapping potential. The population of the dressed molecules are given as \cite{bcsbecwy}
\begin{equation}
\tilde{n}_b(\mathbf{\tilde{r}})=\langle d_0^{\dag}d_0\rangle=
|\Delta(\mathbf{\tilde{r}})|^2
\left[\alpha_p-\frac{U_p(\delta_p-2\mu(\mathbf{\tilde{r}}))}{\alpha_p}\right]^{-2}.
\end{equation}
\begin{figure}[tbp]
\centering
\includegraphics[width=4.5cm]{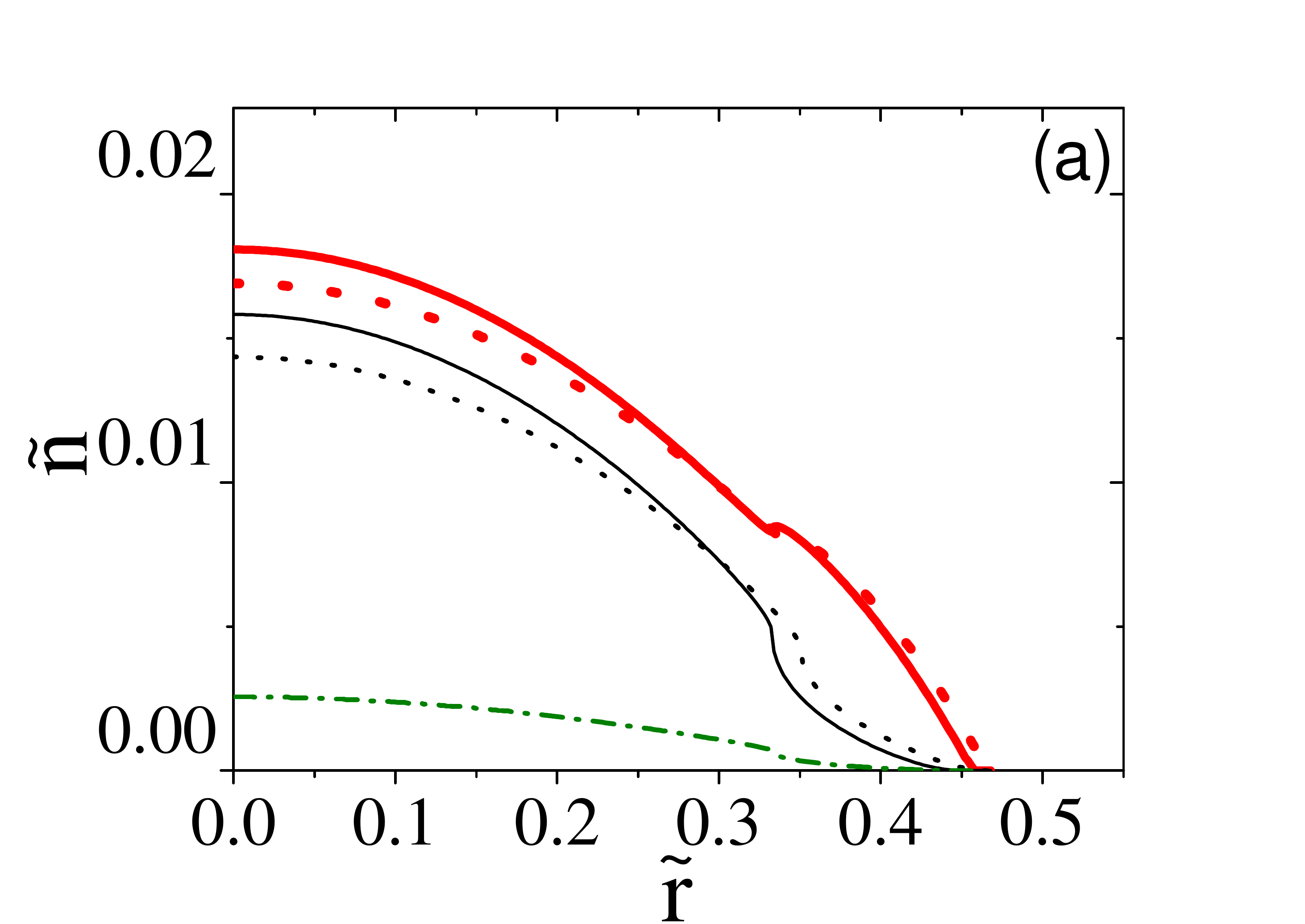}
\hskip-0.5cm
\includegraphics[width=4.5cm]{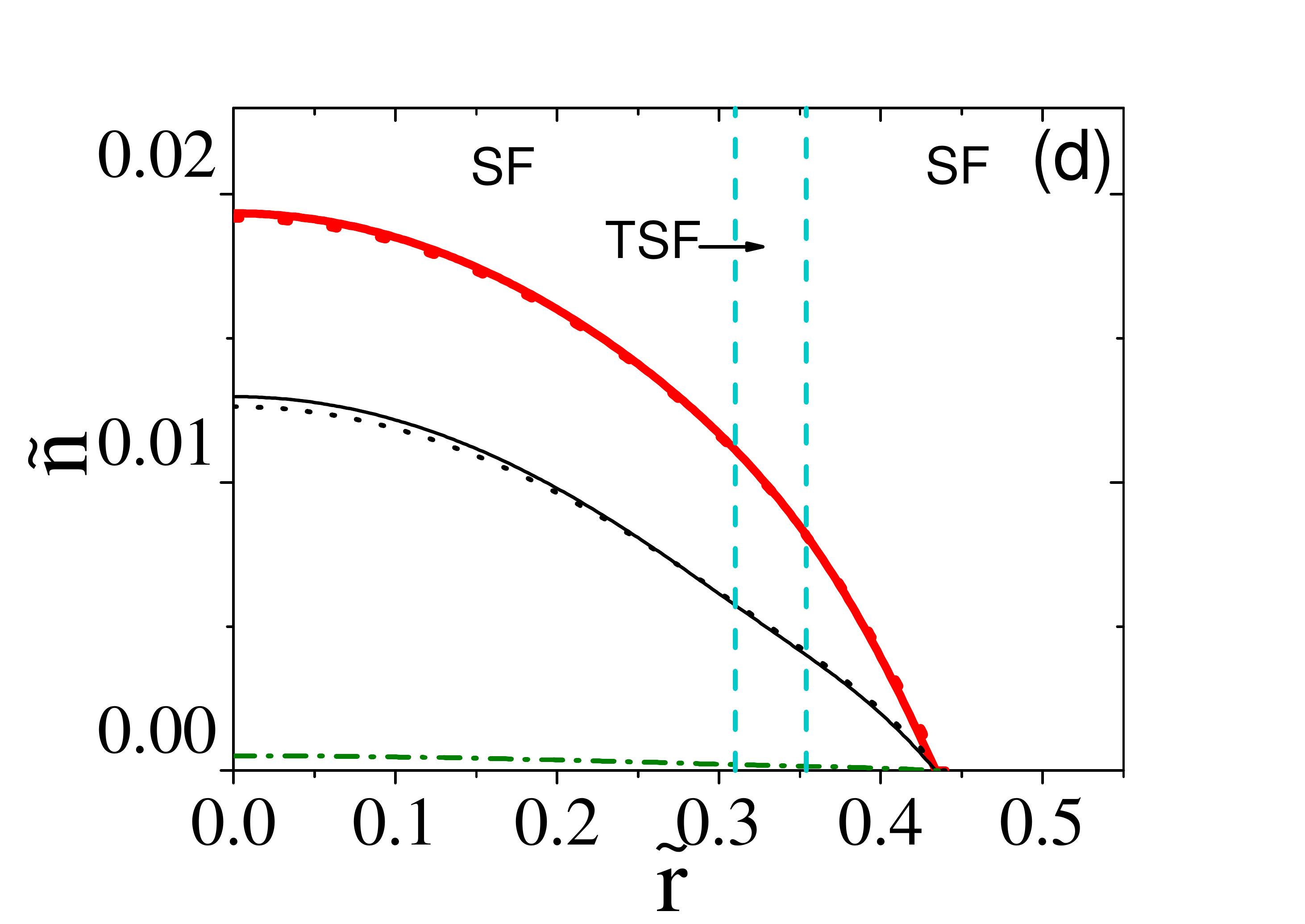}
\includegraphics[width=4.5cm]{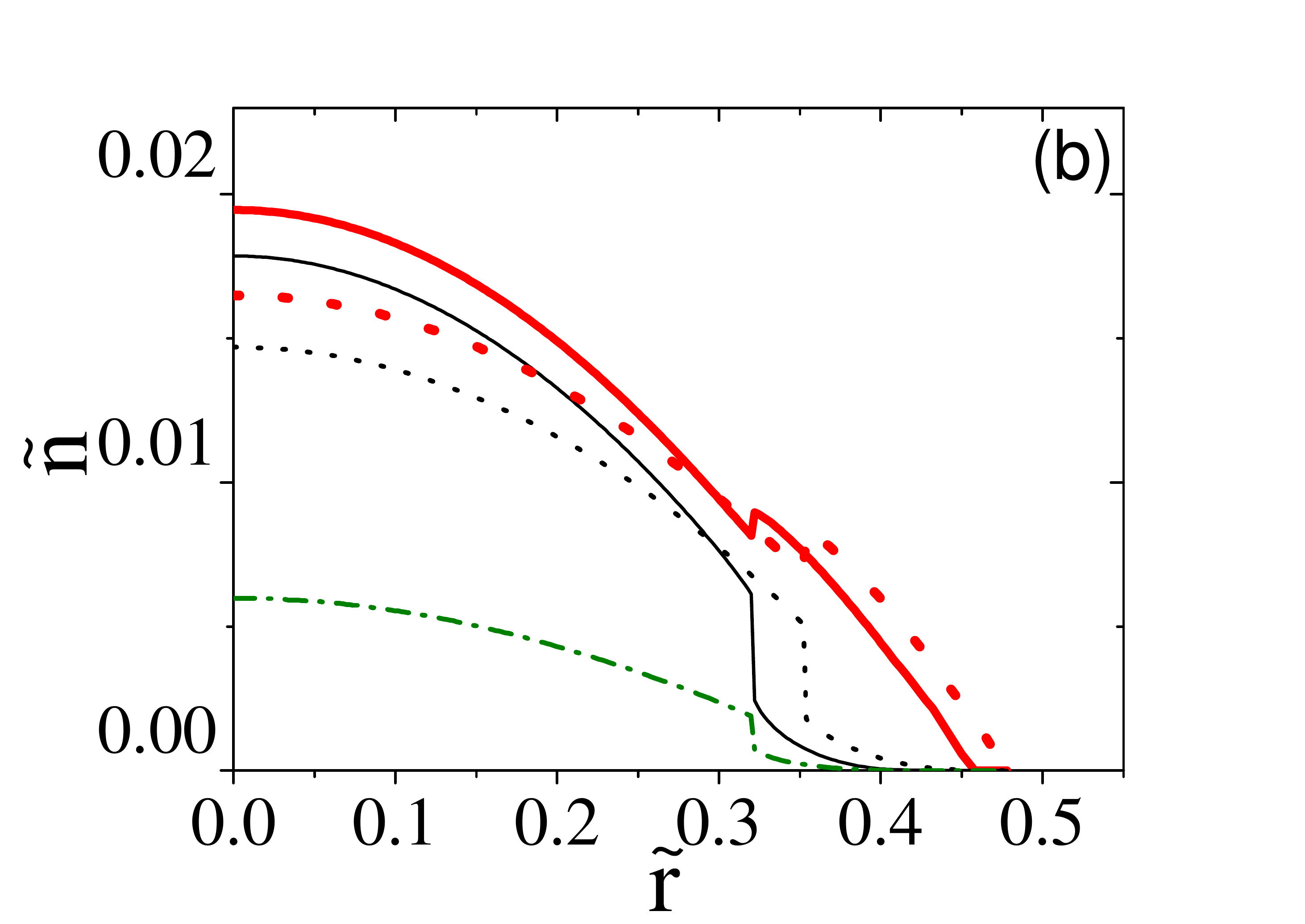}
\hskip-0.5cm
\includegraphics[width=4.5cm]{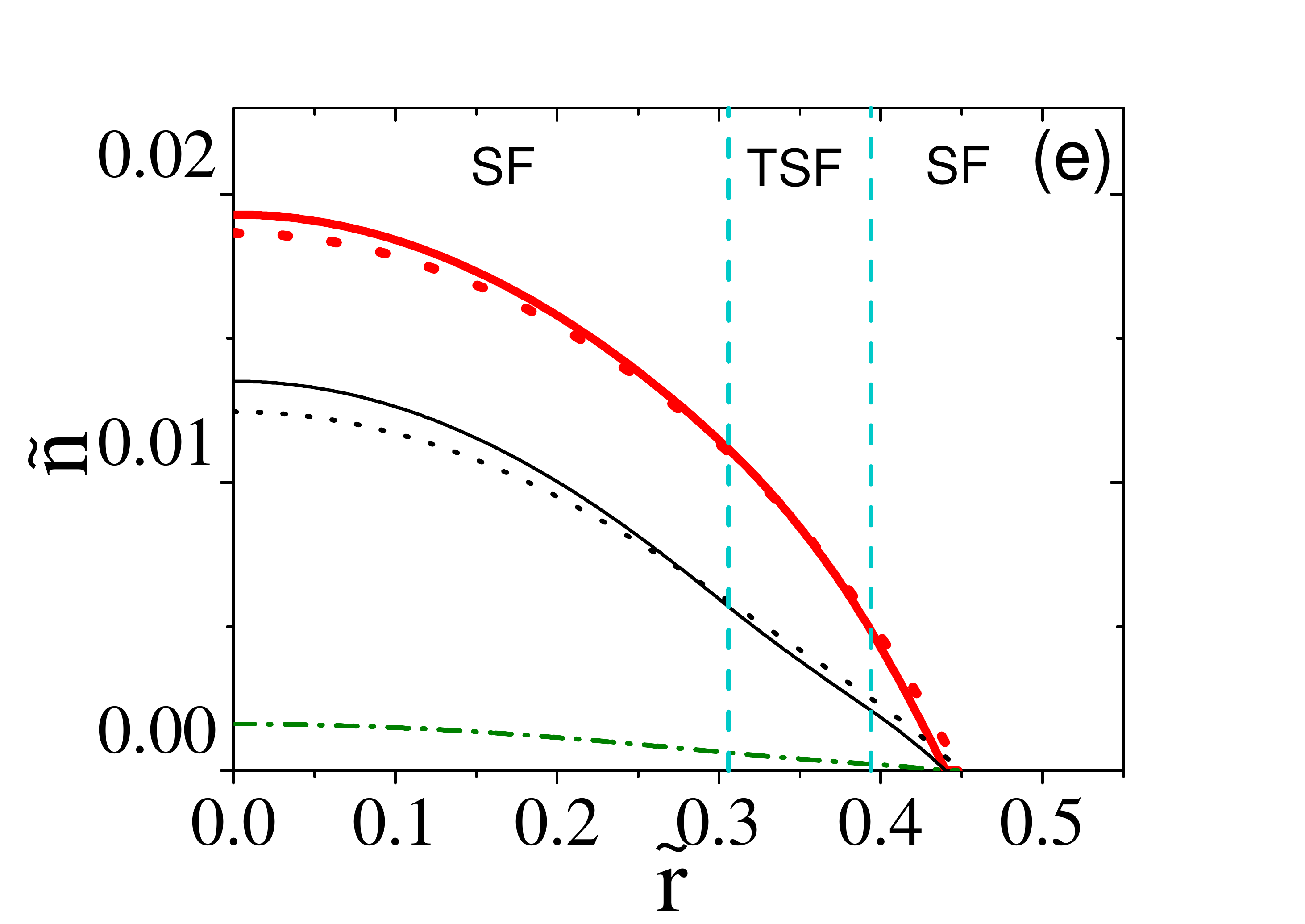}
\includegraphics[width=4.5cm]{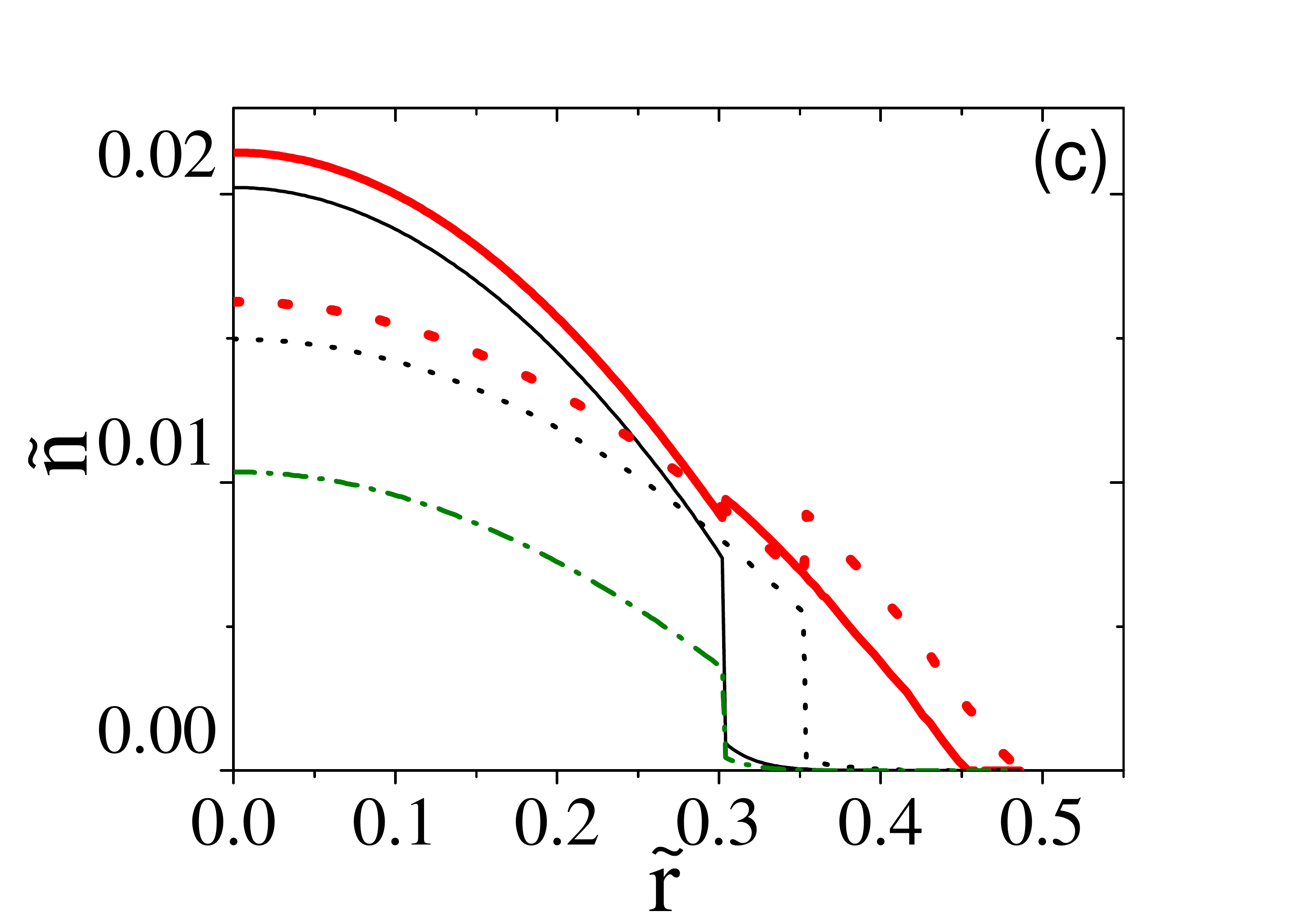}
\hskip-0.5cm
\includegraphics[width=4.5cm]{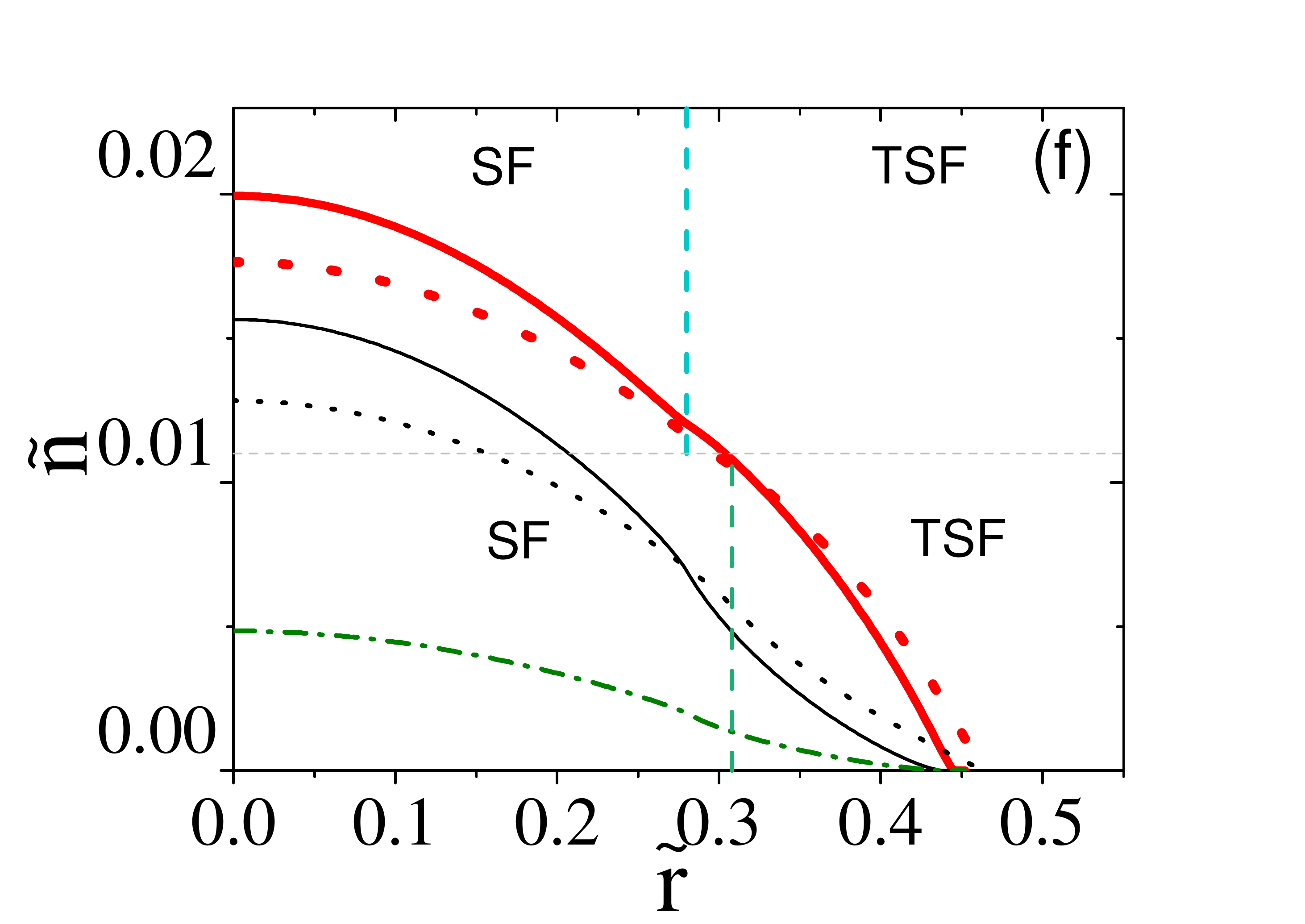}
\caption{(Color online) Density distribution for different interaction strengths with fixed SOC and polarization: (a-c) $\lambda\sim 0.14$, $P\sim 0.20$; (d-f) $\lambda\sim 0.22$, $P\sim 0.27$. In all cases, $\omega_x/\omega_z=10^{-3}$, $N\sim 10^4$, with other parameters: (a) $a_t/a_s\sim -0.36$, $2N_b/N\sim 6\%$; (b) $a_t/a_s\sim 0.04$, $2N_b/N\sim 13\%$; (c) $a_t/a_s\sim 0.34$, $2N_b/N\sim 20\%$; (d) $a_t/a_s\sim -1$, $2N_b/N\sim 1\%$; (e) $a_t/a_s\sim -0.5$, $2N_b/N\sim 4\%$; (f) $a_t/a_s\sim 0$, $2N_b/N\sim 10\%$.
The bold red (thin black) dotted curves are the density distribution of spin-up (spin-down) atoms in the single-channel model, and the solid curves are the density distribution in the effective two-channel model. The distributions for atoms in the closed channel are shown with the dash-dotted lines (green). In (a-c), the phase boundary between the SF and the TSF phases are first-order for both models. While in (d)(e), the vertical lines indicate the second-order phase boundaries between the SF and TSF phases in the two-channel model. In (f), as the TSF phase shows up in both models, we indicate the second-order phase boundary between the SF and TSF phases for the two-channel (single-channel) model using a vertical dashed line in the upper (lower) plane.}\label{figdenas}
\end{figure}

The order parameter characterizing the topological superfluid (TSF) state is the same as that of the superfluid (SF) state, since the two states have the same symmetry and are not separated by a spontaneous symmetry breaking. It has been shown that when the Zeeman field $h$ crosses $\sqrt{\mu^2+\Delta^2}$ from below, an excitation gap closes and then opens up again, while the pairing order parameter remains finite \cite{sau, jzhou}. The system then undergoes a topological phase transition from an SF state to a TSF state \cite{sau}, where the winding number (Chern's number) becomes non-zero and Majorana zero modes can be stabilized at the center of vortex excitations \cite{sau}. Hence, the TSF phase can be identified in the trap where $h>\sqrt{\mu^2(\mathbf{\tilde{r}})+|\Delta(\mathbf{\tilde{r}})|^2}$~\cite{sau,jzhou}.

We now calculate the typical density distributions of a polarized quasi-2D Fermi gas with SOC. In the following discussion, we consider $^{6}$Li as a specific example, while results for $^{40}$K are similar on a qualitative level.
To see the effect of dressed molecules in the closed channel, we compare results from the effective two-channel model with those of a single-channel model \cite{jzhou}. In the context of quasi-2D gases, the single-channel model can be understood as an approximation in which the population of all excited levels in the axial direction are neglected. One may then integrate out the lowest axial mode and relate the two-body binding energy $|E_b|$ appearing in the renormalization condition in the single-channel model to the 3D scattering length $a_s$ via $|E_b| \approx 0.915 \pi^{-1} \exp\left(\sqrt{2\pi}a_t/a_s\right) \hbar\omega_z$ \cite{randeria,petrov,wouters}.
\begin{figure*}[tbp]
\centering
\includegraphics[width=5cm]{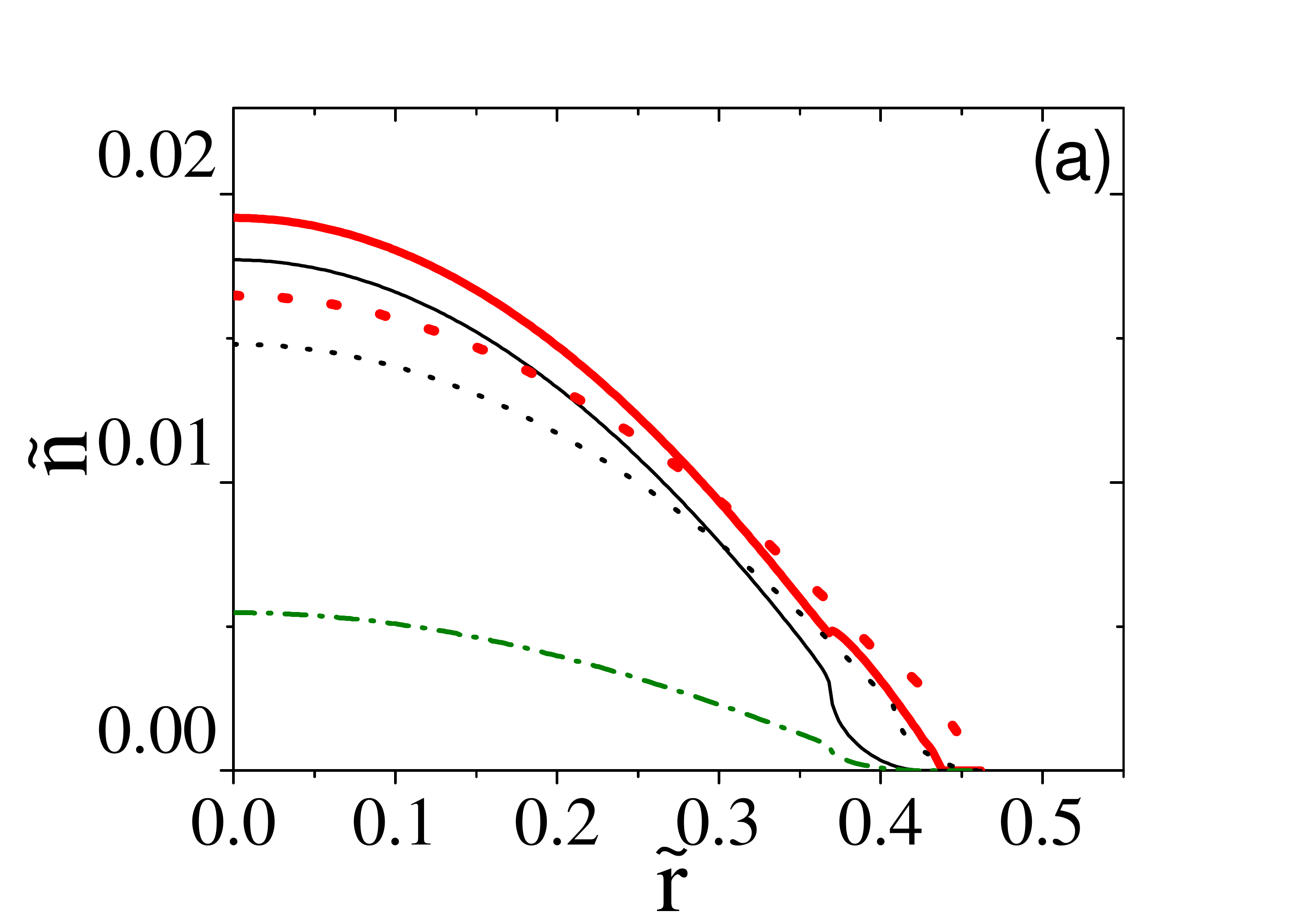}
\includegraphics[width=5cm]{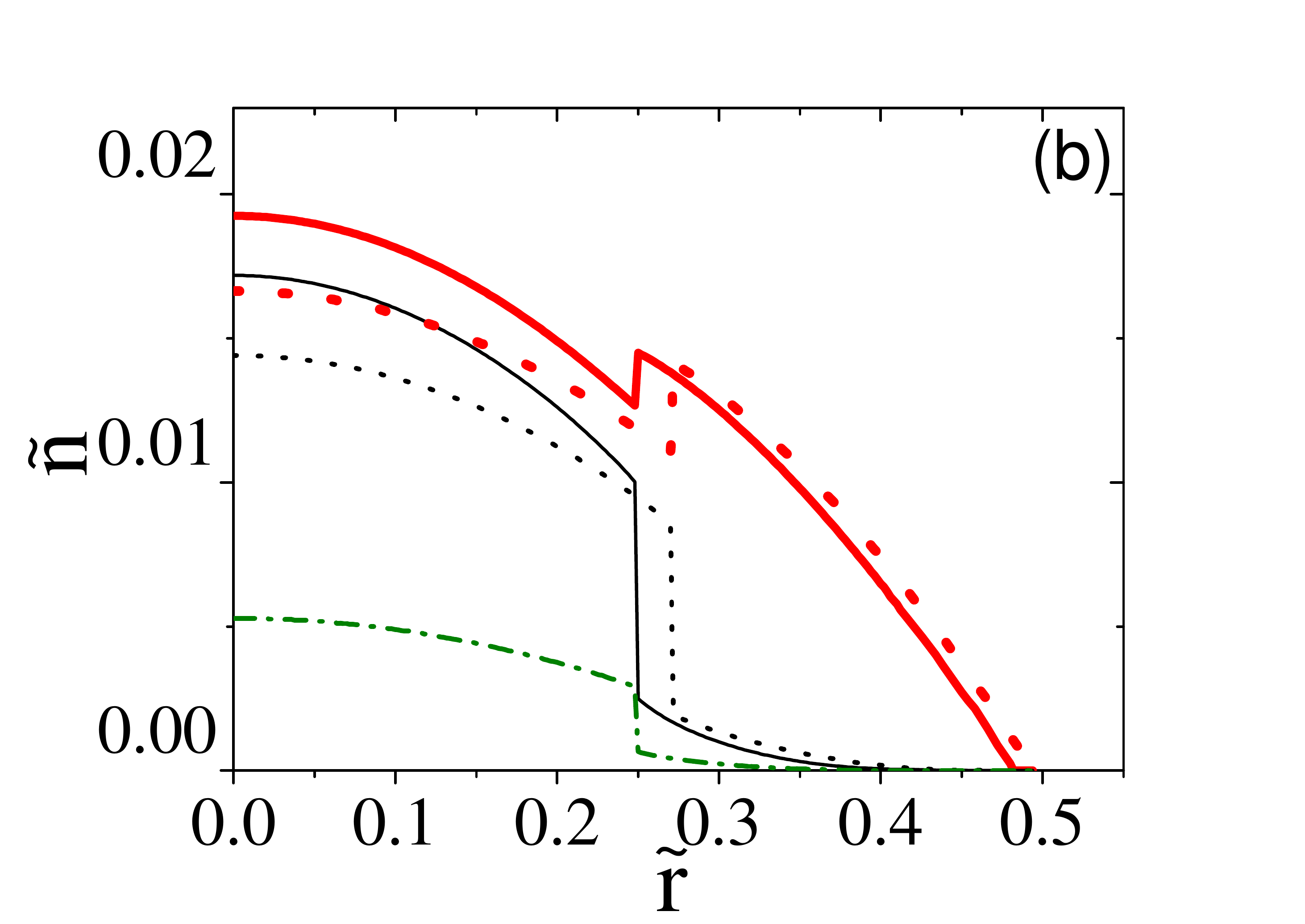}
\includegraphics[width=5cm]{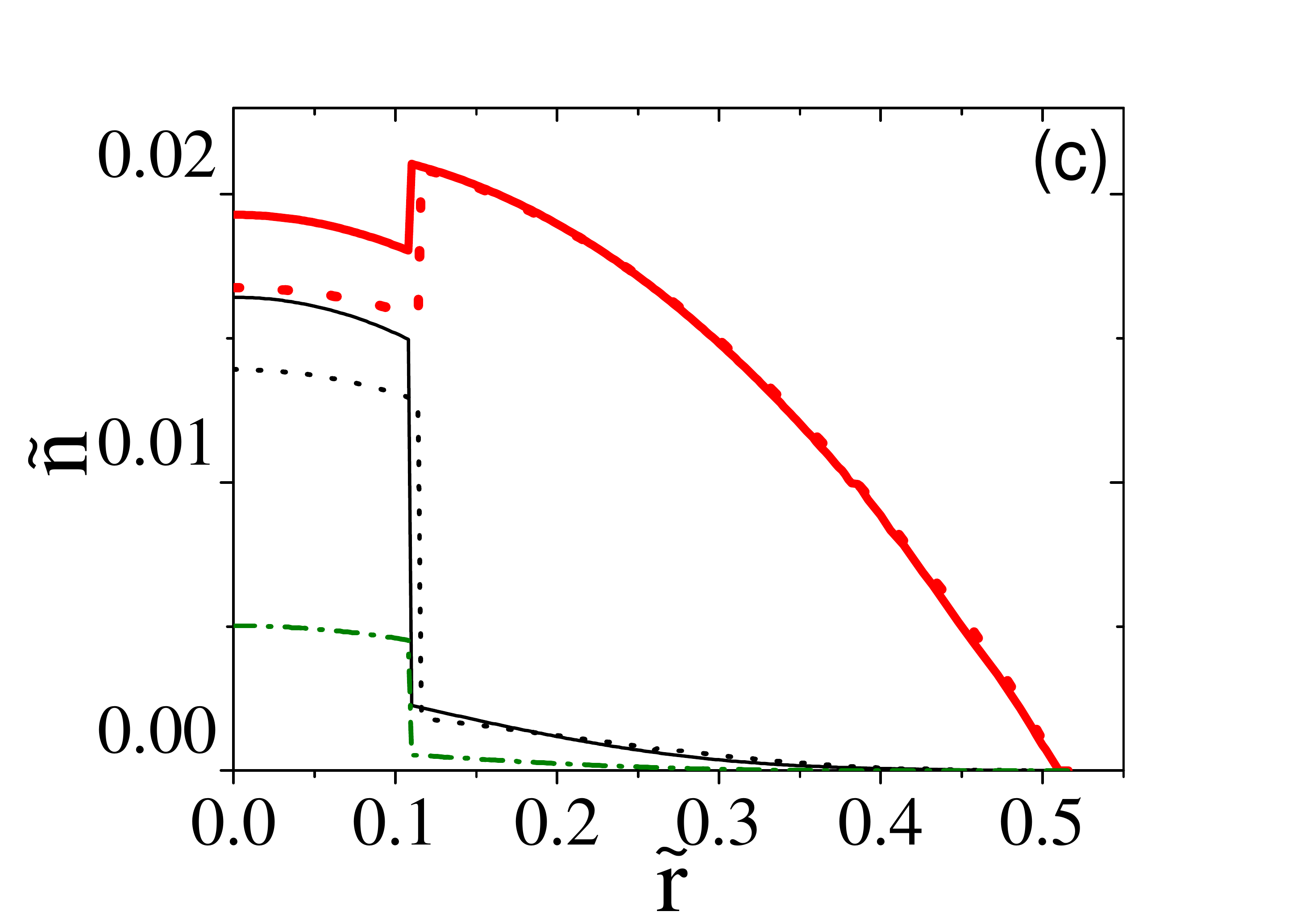}
\caption{(Color online) Density distribution for different polarization at resonance $a_t/a_s=0$. In all cases, $\lambda\sim 0.14$, $\omega_x/\omega_z=10^{-3}$, $N\sim 10^4$, with the polarizations and the closed-channel populations: (a) $P\sim 0.10$, $2N_b/N\sim 13\%$; (b) $P\sim 0.42$, $2N_b/N\sim 8\%$; (c) $P\sim 0.82$, $2N_b/N\sim 2\%$. The bold red (thin black) dotted curves are the density distribution of spin-up (spin-down) atoms in the single-channel model, and the solid curves are the density distribution in the effective two-channel model. The distributions for atoms in the closed channel are shown with the dash-dotted lines (green). Only first-order phase boundaries appear here for both models.}\label{figdenP}
\end{figure*}

In Fig.~\ref{figden1} we demonstrate the typical density distributions calculated from the two models. From these results, we find that the dressed molecules affect properties of the trapped gas in two different ways.
First, the density distribution can be significantly modified by the inclusion of dressed molecules [c.f. Fig.~\ref{figden1}(a)]. Second, in the presence of dressed molecules, the in-trap phase structure can be different. This is manifested in Fig.~\ref{figden1}(a), where the first-order phase boundaries between the conventional superfluid phase (SF) at the center and the topological superfluid phase (TSF) at the edge are shifted. Notably, on the BCS side of the Feshbach resonance, the dressed molecules can qualitatively alter the in-trap phase structure by inducing an additional TSF phase which is absent in a single-channel calculation [c.f. Fig.~\ref{figden1}(b)].

In Fig.~\ref{figdenas}, we systematically map out the density distributions for fixed total particle number and polarization while the interaction strength is tuned. Apparently, the difference in density distribution becomes more dramatic as the interaction strength is tuned toward the BEC side. This also applies to the shift in the phase boundary between the SF and TSF phases [Fig.~\ref{figdenas} (a-c)]. On the other hand, the stability region of the TSF phase induced by the dressed molecules increases toward the BEC side [Fig.~\ref{figdenas}(d-f)]. The TSF phase in the single-channel model appears near unitarity, after which the two models differ only by a shift in the phase boundaries [Fig.~\ref{figdenas}(f)]. Importantly, in either case, the inclusion of dressed molecules results in significant changes to the many-body properties of the system that may be probed experimentally.

The effects of dressed molecules on the density distribution in quasi-2D Fermi gases have been studied before for an
unpolarized case without SOC~\cite{wzduan,wzduan2}. In that case, the modification of the density
distribution is only significant in the deep BEC region. In the current case, however, the dressed molecules
can have measurable effects near unitarity or even on the BCS side of the resonance.
This is a combined effect of SOC and population imbalance. First, as discussed in the two-body calculation,
SOC causes an increase of two-body binding energy, which can significantly populate the excited axial harmonic states and consequently enhance the effects of dressed molecules.
Second, it is also related to the mechanism by which the population imbalance is accommodated in the system.
Generally speaking, in a trapped Fermi gas, the population imbalance is accounted for by the normal phase
toward the trap edge or by exotic superfluid phases that supports polarization,
e.g. the FFLO phase~\cite{fflo1,fflo2}, the BP phase~\cite{bp1},
or the TSF phase in the presence of SOC~\cite{zhang-sarma,sato}. The dressed molecules,
on the other hand, do not support polarization. Therefore, when the dressed molecules are present,
the in-trap density distribution must be modified to accommodate the total polarization fixed {\it a priori}.
A particularly interesting observation here is that exotic phases may be induced in the trap to account for
polarization even in the presence of very small dressed molecule populations [Fig.~\ref{figdenas}(d-e)].
We expect that this picture should hold for a quasi-2D polarized Fermi gas {\it without} SOC.

Finally, we plot in Fig.~\ref{figdenP} density distributions for various polarizations at resonance.
As polarization increases, the population of the dressed molecules decreases. This can be understood
in terms of the competition between polarization and closed-channel population,
as larger polarization requires more population in the open channel. The results suggest that at resonance, the effects of dressed molecules are stronger at intermediate polarization.

\section{Summary}

We study a quasi-two-dimensional Fermi gas confined in a strong axial harmonic
potential with Rashba spin-orbit coupling and population imbalance. We analyze the
two-body bound state in such a system, and find that the axial excited harmonic levels
can be significantly populated as the binding energy becomes comparable to or exceeds the
axial confinement. The presence of Rashba spin-orbit coupling increases the density of
states in the low-energy limit, hence causes an enhancement of two-body binding energy,
and eventually excites more particles to the high-lying axial harmonic modes.
This observation implies that the degrees of freedom along the third dimension would be
important for a satisfactory description of the quasi-two-dimensional,
but truly three-dimensional system.

In order to incorporate the effects of these axial excited modes, we introduce a dressed
molecule degree of freedom and construct an effective two-dimensional Hamiltonian
in the form of a two-channel model. The parameters in this effective model is fixed by matching single- and two-body physics
with the original Hamiltonian. Specifically, we require the effective Hamiltonian to give the correct single-particle dispersion,
background scattering amplitude, two-body binding energy, and fraction of Feshbach molecules plus dimers formed with axial excited fermions.
These matching conditions are equivalent to fixing the pole of two-body $T$-matrix and the first derivative of
$T^{-1}$ around the pole. Therefore, the effective model can mimic the low-energy physics as the original Hamiltonian
provided that the chemical potential is not too far away from one half of the two-body binding energy.
This condition usually holds for ultra-cold dilute Fermi gases in quasi-2D confinement.

We then investigate many-body properties of a quasi-2D Fermi gas using this effective Hamiltonian.
We conclude that the inclusion of dressed molecules are crucial for the investigation of the underlying
system. Specifically, we find that the stability region of the topological superfluid phase is increased as a result
of the occupation of axial excited modes. We then systematically map out the in-trap phase structures and
density distributions throughout the whole BCS-BEC crossover region, and discuss the appropriate parameter region that the effects of the dressed molecules may be observed. Our findings are helpful for the experimental
search for the topological superfluid phase in ultra-cold Fermi gases, and have interesting implications for quasi-low-dimensional polarized Fermi gases in general.
\acknowledgments
This work is supported by NKBRP (2013CB922000), NFRP (2011CB921200), NNSF (60921091), NSFC (10904172, 11105134, 11274009), the Fundamental Research Funds for the Central Universities (WK2470000001, WK2470000006), and the Research Funds of Renmin University of China (10XNL016, 13XNH123). F.W. acknowledges the support by the Fund for Fostering Talents in Basic Science of the National Natural Science Foundation of China (No.J1103207).  W.Z. would also like to thank the NCET program for support.


\begin{thebibliography}{99}

\bibitem{lin-exp2} Y.-J. Lin, K. Jim\'{e}nez-Garc\'{i}a, I. B. Spielman,
Nature(London) {\bf 471}, 83 (2011).

\bibitem{zhang-exp} P. Wang, Z. Q. Yu, Z. Fu, J. Miao, L. Huang, S. Chai, H. Zhai, J. Zhang,
 Phys. Rev. Lett. {\bf 109}, 095301 (2012).

\bibitem{mit-exp} L. W. Cheuk, A. T. Sommer, Z. Hadzibabic, T. Yefsah, W. S. Bakr, M. W. Zwierlein,
Phys. Rev. Lett. {\bf 109}, 095302 (2012).

\bibitem{zhang-sarma} C. Zhang, S. Tewari, R. M. Lutchyn, S. Das Sarma,
Phys. Rev. Lett. {\bf 101}, 160401 (2008).

\bibitem{sato} M. Sato, Y. Takahashi, S. Fujimoto, Phys. Rev. Lett. \textbf{103}, 020401 (2009).

\bibitem{shenoy} J. P. Vyasanakere, V. B. Shenoy,  Phys. Rev. B {\bf 83}, 094515 (2011).

\bibitem{gongming} M. Gong, S. Tewari, C. Zhang,  Phys. Rev. Lett. {\bf 107}, 195303 (2011).

\bibitem{yu} Z.-Q. Yu, H. Zhai, Phys. Rev. Lett. {\bf 107}, 195305 (2011).

\bibitem{hu} H. Hu, L. Jiang, X. J. Liu, H. Pu, Phys. Rev. Lett. {\bf 107}, 195304 (2011).

\bibitem{iskin} M. Iskin, A. L. Subasi, Phys. Rev. Lett. {\bf 107}, 195304 (2011).

\bibitem{yi} W. Yi, G.-C. Guo, Phys. Rev. A {\bf 84}, 031608(R) (2011).

\bibitem{sademelo} L. Han, C. A. R. S{\'a} de Melo,  Phys. Rev. A {\bf 85}, 011606(R) (2012).

\bibitem{jzhou} J. Zhou, W. Zhang, W. Yi, Phys. Rev. A \textbf{84}, 063603 (2011).

\bibitem{yang} X. Yang, S. Wan,  Phys. Rev. A {\bf 85}, 023633 (2012).

\bibitem{chuanwei} M. Gong, G. Chen, S. Jia, C. Zhang,
 Phys. Rev. Lett. {\bf 109}, 105302 (2012).

\bibitem{helianyi} L. He, X. G. Huang,
Phys. Rev. Lett. {\bf 108}, 145302 (2012).

\bibitem{sau} J. D. Sau, R. M. Lutchyn, S. Tewari, S. Das Sarma, Phys. Rev. Lett. {\bf 104}, 040502 (2010).

\bibitem{ol} M. K\"ohl, H. Moritz, T. Stoferle, K. Gunter, T. Esslinger, Phys. Rev. Lett. \textbf{94}, 080403 (2005).

\bibitem{vale} P. Dyke, E. D. Kuhnle, S. Whitlock, H. Hu, M. Mark, S. Hoinka, M. Lingham, P. Hannaford, C. J. Vale, Phys. Rev. Lett. {\bf 106}, 105304 (2011).

\bibitem{randeria} M. Randeria, J. M. Duan, L. Y. Shieh, Phys. Rev. B {\bf 41}, 327 (1990).

\bibitem{petrov} D. S. Petrov, G. V. Shlyapnikov, Phys. Rev. A \textbf{64}, 012706 (2001).

\bibitem{wouters} J. Tempere, M. Wouters, J. T. Devreese, Phys. Rev. B \textbf{75}, 184526 (2007).

\bibitem{jason-pra06} J. P. Kestner, L.-M. Duan, Phys. Rev. A \textbf{74}, 053606 (2006).

\bibitem{Duan-07} J. P. Kestner, L.-M. Duan,  Phys. Rev. A \textbf{76}, 063610 (2007).

\bibitem{wzduan} W. Zhang, G.-D. Lin, L.-M. Duan, Phys. Rev. A \textbf{77}, 063613 (2008).

\bibitem{wzduan2} W. Zhang, G.-D. Lin, L.-M. Duan, Phys. Rev. A \textbf{78}, 043617 (2008).

\bibitem{demler} V. Pietil{\"a}, D. Pekker, Y. Nishida, E. Demler, Phys. Rev. A {\bf 85}, 023621 (2012).

\bibitem{fflo1} P. Fulde, R. A. Ferrell, Phys. Rev. \textbf{135}, A550 (1964).

\bibitem{fflo2} A. I. Larkin, Y. N. Ovchinnikov, Sov. Phys. JETP \textbf{20}, 762 (1965).

\bibitem{bp1} W. V. Liu, F. Wilczek, Phys. Rev. Lett. \textbf{90}, 047002 (2003).

\bibitem{2dgasexp1} K. Martiyanov, V. Makhalov, A. Turlapov, Phys. Rev. Lett. {\bf 105}, 030404 (2010).

\bibitem{2dgasexp2} B. Fr\"ohlich, M. Feld, E. Vogt~E, M. Koschorreck, W. Zwerger, M. K\"ohl, Phys. Rev. Lett. {\bf 106}, 105301 (2011).

\bibitem{2dgasexp3} A. T. Sommer, L. W. Cheuk, M. J. H. Ku, W. S. Bakr, M. W. Zwierlein, Phys. Rev. Lett. {\bf 108}, 045302 (2012).

\bibitem{2dgasexp4} Y. Zhang, W. Ong, I. Arakelyan, J. E. Thomas~J~E, Phys. Rev. Lett. {\bf 108}, 235302 (2012).

\bibitem{holland-01} M. Holland, S. J. J. M. F. Kokkelmans, M. L. Chiofalo, R. Walser, Phys. Rev. Lett. \textbf{87}, 120406 (2001).

\bibitem{timmermans-99} E. Timmermans, P. Tommasini, M. Hussein, A. Kerman, Phys. Rep. \textbf{315}, 199 (1999).

\bibitem{qijinpr} Q. Chen, J. Stajic, S. Tan, K. Levin, Phys. Rep. \textbf{412}, 1 (2005).

\bibitem{kresonance1} C. A. Regal, M. Greiner, D. S. Jin, Phys. Rev. Lett \textbf{92}, 040403 (2004).

\bibitem{kresonance2} M. W. Zwierlein, C. A. Stan, C. H. Schunck, S. M. F. Raupach, A. J. Kerman, W. Ketterle, Phys. Rev. Lett. \textbf{92}, 120403 (2004).

\bibitem{kresonance3} C. Chin, M. Bartenstein, A. Altmeyer, S. Riedl, S. Jochim, J. Hecker Denschlag, R. Grimm, Science \textbf{305}, 1128 (2004).

\bibitem{liresonance} M. Bartenstein, A. Altmeyer, S. Riedl, R. Geursen, S. Jochim, C. Chin, J. Hecker Denschlag,
R. Grimm, A. Simoni, E. Tiesinga, C. J. Williams, P. S. Julienne, Phys. Rev. Lett. \textbf{94}, 103201 (2005).

\bibitem{bcsbecwy} W. Yi, L.-M. Duan, Phys. Rev. A \textbf{73}, 063607 (2006).
\end{thebibliography}
\end{document}